\documentclass[article, pra, showpacs, twocolumn, showkeys, secnumarabic, aps, amsmath, amssymb, nofootinbib, superscriptaddress, longbibliography, floatfix, table-of-contents, dblfloatfix]{revtex4-2}

\usepackage[utf8]{inputenc}
\usepackage[pdftex]{graphicx}
\usepackage{caption}
\usepackage{subcaption}
\usepackage{mathrsfs}
\usepackage[colorlinks, breaklinks, urlcolor={blue}, linkcolor={red}, citecolor={blue}]{hyperref}
\usepackage{array}
\usepackage{amsmath}
\usepackage{type1cm}
\usepackage[export]{adjustbox}
\usepackage{dsfont}
\usepackage{lettrine}
\usepackage[english]{babel}
\usepackage{lmodern}
\usepackage{microtype}
\usepackage{booktabs}
\usepackage[T1]{fontenc}
\usepackage[boxed, vlined]{algorithm2e}
\usepackage{braket}
\usepackage{xcolor}
\usepackage{orcidlink}
\usepackage{braket}
\usepackage{bm}
\usepackage{bbold}
\usepackage{tikz}

\usepackage{upgreek}
\DeclareMathOperator{\Tr}{Tr}

\begin{document}
	
	\title{Enhancing the Charging Performance of Many-Body Quantum Batteries through Landau-Zener Driving}

	\author{Syed Abubacker Siddique}
	\affiliation{Department of Physics and Nanotechnology, Faculty of Engineering and Technology, SRM Institute of Science and Technology, Chennai - 603203, Tamil Nadu, India}
	
	\author{Md~Manirul Ali  \orcidlink{0000-0002-5076-7619}}
	\email[]{manirul@citchennai.net}
	\affiliation{Centre for Quantum Science and Technology, Chennai Institute of Technology, Chennai - 600069, India}
	
	\author{Arijit Sen \orcidlink{0000-0002-8624-9418}}
	\email[]{arijits@srmist.edu.in}
	\affiliation{Department of Physics and Nanotechnology, Faculty of Engineering and Technology, SRM Institute of Science and Technology, Chennai - 603203, Tamil Nadu, India}
	
	\date{\today}
	
\begin{abstract}
We explore the charging advantages of a many-body quantum battery driven by a Landau-Zener field. Such a system may be modeled as a Heisenberg XY spin chain with \textit{N} interacting spin-$\frac{1}{2}$ particles under an external magnetic field. Here we consider both nearest-neighbor and long-range spin interactions. The charging performance of this many-body quantum battery is evaluated by comparing Landau-Zener and periodic driving protocols within these interaction regimes. Our findings show that the Landau-Zener driving can offer superior energy deposition and storage efficiency compared to periodic driving. Notably, the Landau-Zener protocol may deliver optimal performance when combined with long-range interactions. The efficiency of a Landau-Zener quantum battery can be significantly enhanced by optimizing key parameters, such as XY anisotropy, the magnitude of the driving field, and interaction strength.
\end{abstract}
	
	
\maketitle
	
\section{Introduction}
	
\noindent
	
Developments in quantum technologies have led to applications such as quantum computation \cite{nielsen2010quantum}, quantum communication \cite{xu2020secure}, quantum energy storage \cite{campaioli2018quantum,quach2023quantum,campaioli2024colloquium}, quantum
meterology and sensing \cite{xiang2013quantum}. As devices continue to shrink in size, quantum effects are playing an increasingly significant role in the advancing field of quantum technology. Energy can be temporarily stored in miniature devices using quantum batteries. The concept of the quantum battery was first introduced in the seminal work of Alicki and Fannes \cite{alicki2013entanglement}.This breakthrough opened new avenues for researchers to explore quantum phenomena, enabling the use of quantum resources to enhance battery functionality and achieve high-performance energy manipulation \cite{ferraro2018high,rossini2020quantum,xu2021enhancing,seah2021quantum,campaioli2017enhancing,ghosh2020enhancement,andolina2019extractable}. There is a growing interest in utilizing quantum systems as heat engines \cite{bhattacharjee2021quantum,cangemi2024quantum} and energy storage devices \cite{campaioli2018quantum,quach2023quantum,campaioli2024colloquium}. Understanding their operation under various charging conditions thus remains a topic of significant focus at present.
	
In the simplest case, where environmental dissipation can be neglected, a unitary protocol is used to charge or transfer energy to the battery. A major advancement in this field occurred with the proposal of the Dicke model for a quantum battery, where \textit{N} two-level systems are coupled to a single photonic mode in a cavity \cite{ferraro2018high}. Since then, numerous significant quantum battery models have been theoretically proposed, ranging from one-dimensional chains\cite{le2018spin} to strongly interacting SYK fermionic batteries\cite{rossini2020quantum}. The charging or energy transfer to the battery is achieved through a driven unitary transformation\cite{hadipour2023study}. Given that the quantum system begins in the ground state of its Hamiltonian, the charging process involves turning ON the external driving field for a finite duration, during which the battery evolves unitarily, driven by a net ``charging Hamiltonian''. Towards the end of the charging protocol, the external driving fields are turned OFF, and the energy difference between the final and initial states, with respect to the battery's Hamiltonian, is considered as the energy deposited in the battery\cite{vsafranek2023work}.
	
However, in the reverse process, it is typically not possible to extract all of the energy stored in the battery using unitary protocols. The maximum amount of useful energy that can be extracted is referred to as \textit{ergotropy}. The rate at which energy is stored in the battery is called the \textit{charging power}, and ideally, it should also be maximized\cite{ferraro2018high}. It is also important to note that a high variance in the charged state of the battery with respect to the battery Hamiltonian indicates undesirable instability in the energized state of the battery\cite{bhattacharjee2021quantum}. When considering a battery composed of \textit{N} identical quantum cells, each operating as an independent system, parallel charging approaches that charge each cell separately do not lead to an increase in charging power proportional to \textit{N} \cite{campaioli2017enhancing}. However, if a coherent charging protocol is employed, where the charging Hamiltonian induces interactions between multiple cells, it can lead to a superlinear increase in charging power with the system size. This phenomenon is often referred to as a \textit{quantum advantage}\cite{ferraro2018high}. In this paper, we slightly step aside from the conventional approach of charging a quantum battery using a time-periodic driving protocol and instead  adopt a linear driving protocol. In other words, we consider a charging Hamiltonian that is linear in time and analyze the energetics of the battery at an arbitrary time interval, which corresponds to the duration of the time period in the charging Hamiltonian\cite{mitchison2021charging,downing2023quantum}. As we will observe, this approach gives rise to distinct and useful behavior in the quantum battery. While the concept of a quantum battery was not initially associated with a many-body quantum system, the charging process was originally framed as a global one – that is, the state evolves within the total Hilbert space of \textit{N} batteries. In contrast, we consider the case of a many-body quantum battery, where the intrinsic two-body interactions between the system’s constituents can lead to entanglement. This allows charging to occur in one part of the battery, while the entire system approaches a collective excited state, ultimately achieving the maximum average energy\cite{campaioli2018quantum}. Quantum batteries can be realized in various ways\cite{liu2019loss,catalano2023frustrating}. Advances in trapping ultracold atoms with light patterns (optical lattices), ions in traps, and even polar molecules have made it feasible to build and manipulate the core components of quantum batteries - complex systems governed by quantum mechanics. This progress paves the way for developing practical quantum technologies based on these approaches\cite{jin2011polar,lewenstein2007ultracold}.

Here, we consider a simple model based on a spin lattice configuration of \textit{N} interacting half-integer spins. The ground state of a spin model with interactions can serve as the initial state of the battery. The battery functions as a storage device, with its charging facilitated through quantum-mechanically permissible operations\cite{andolina2018charger}. This approach has been extensively studied in recent years and has gained further exploration in recent works\cite{grazi2024enhancing,du2024quantum}. In this work, we focus on a scenario where the initial state of the battery is the ground state of the quantum spin chain, and a time-dependent charging field is applied to drive the system, aiming to extract power from the battery. This paper is structured as follows: In Sec. II, we present our spin-chain battery model, considering both nearest-neighbor and long-range interactions. In Sec. III, we explore the role of a time-dependent external charging field and the charging mechanism to assess how effectively Landau-Zener driving influences the energetics of a many-body interacting system. In Sec. IV, we discuss the results on work deposition and the average power transferred under different scenarios, while varying the anisotropy, Landau-Zener driving parameter, and interaction strength. In Sec. V, we compare our results for Landau-Zener driving with those obtained using a time-periodic driving field, and examine how work deposition changes as the number of spins increases. Finally, we conclude in Sec. VI.
\newline

\section{Many-body Quantum battery model}
	
	We investigate a many-body quantum battery, represented by the Hamiltonian of a quantum $XY$ Heisenberg spin chain, which consists of $N$ interacting spin-1/2 systems in the presence of a magnetic field. The spin chain has the potential to be used in other quantum devices, communications, and as a working systems \cite{zueco2009quantum,marchukov2016quantum}. Experimental evidences provide a wide range of possibilities for exploring spin systems such as CuCl$_2$.DMSO (Dimethyl sulfoxide), CuCl$_2$.TMSO(Tetramethyl sulfoxide), (C$_6$H$_{11}$NH$_3$)CuCl \cite{cullen1983monte}. In absence of charging field, the battery Hamiltonian takes the following form \cite{ghosh2020enhancement}
\begin{eqnarray}
\label{H0}
H_0 = H_1 + H_2
\end{eqnarray}
where $H_{1}$ is the external magnetic field and does the work of splitting the degeneracy between the spins $\ket{\downarrow}$ and $\ket{\uparrow}$.
\begin{widetext}
\begin{eqnarray}
\label{HB}
H_0 = \underbrace{B \sum_{i=1}^{N} \sigma_{i}^{z}}_{\text{H$_{1}$}} \underbrace{ - \frac{1}{2} \sum_{i < j}^N g_{ij}
\Big[ \left(1+\gamma \right) \sigma_{i}^{x} \otimes \sigma_{j}^{x}
+  \left(1-\gamma \right) \sigma_{i}^{y} \otimes \sigma_{j}^{y} \Big]}_{\text{H$_2$}},
\end{eqnarray}
\end{widetext}
where $i$ refers to the $i$th spin in the chain and $\sigma_{i,j}$ represents Pauli spin operators. In $H_2$, the second and third terms in the Hamiltonian define pairwise interactions between different spins \cite{le2018spin}. Interaction strengths between spins $i$ and $j$ are represented by $g_{ij}$. We consider both nearest-neighbour and Long-range interaction. The anisotropy is introduced through the parameter $\gamma$ with $-1\le \gamma \le1$. First, we consider the situation when the interactions are nearest site interaction with the nearest site coupling strength given by
\begin{eqnarray}
\label{NN}
g_{ij}=g_{ij}^{NN} = g \delta_{i, j-1}
\end{eqnarray}
and we take long-range interaction as
\begin{eqnarray}
\label{LR}
g_{ij}=g_{ij}^{LR} = \frac{g}{\vert i - j \vert^a}
\end{eqnarray}
where $a$ is a non-negative number and $g$ is a real constant. While considering the long-range interaction case, we take $a=1$ in Eq \eqref{LR}. We consider the attractive interaction $(g \geq 0)$ in such a way that the ground state of the static Hamiltonian is ferromagnetic. We have taken the ground state of our $H_0$ Hamiltonian to be our initial state of the battery throughout this paper. All the numerical simulations were performed using QuTiP \cite{johansson2012qutip}.

\section{Charging protocol under time-dependent linear driving field}

There are various forms of charging involved in the charging process which include time-dependent \cite{rossini2019many,mondal2022periodically,zhang2019powerful} and time-independent Hamiltonians \cite{dziarmaga2005dynamics,caravelli2021energy}. Here we implement a time-dependent charging
Hamiltonian which is experimentally realizable \cite{bauerle2018coherent}. The explicit time-dependent
driving gives us more control over the charging protocol \cite{mazzoncini2023optimal}. During the charging
process, the state of the battery evolves under the charging Hamiltonian, $H_{c}(\tau)=H_0 + V(\tau)$ for a
duration of time $\tau$. The time-dependent external charging field is given by
\begin{eqnarray}
\label{LZ}
V(\tau) = h(\tau) \sum_{i=1}^{N} \sigma_{i}^{z}
\end{eqnarray}
We consider linear driving, $h(\tau)=v\tau$ in Eq.~(\ref{LZ}), where $v$ is the parameter in the Landau-Zener
Hamiltonian that signify the rate at which the energy difference between the two states of individual spin
changes with time. It essentially measures how quickly the system is driven through the avoided crossing, describing a many-body version of the Landau-Zener Hamiltonian \cite{rubbmark1981dynamical,glasbrenner2023landau}.
The initial state of the battery is taken to be the ground state $\rho(0)$ of the battery Hamiltonian $H_0$. During the charging
process, the time-evolved state of the battery is governed by Liouville-von Neumann equation\footnote{Subsequently, we
have taken $\hbar=1$}
\begin{eqnarray}
\label{von}
\frac{d}{dt} \rho(\tau) = -\frac{i}{\hbar} \left[ H_{c}(\tau),  \rho(\tau) \right].
\end{eqnarray}
The time-evolved state of the battery at the end of the charging process is given by
\begin{eqnarray}
\label{Utau}
\rho(\tau) = U(\tau) \rho(0) U^{\dagger}(\tau),
\end{eqnarray}
where the driven unitary transformation is given by the time-ordered exponential of the charging	Hamiltonian $H_{c}(\tau)$:
\begin{eqnarray}
\label{Expo}
U(\tau) = \mathcal{T} \exp \left[ -i \int_{0}^{\tau} ds~H_{c}(s) \right].
\end{eqnarray}
After we switch off the charging field, the energy stored or the deposited work in the battery is given by
\begin{eqnarray}
\label{work}
W(\tau) = \Tr \left[ \rho(\tau) H_0   \right] - \Tr \left[ \rho(0) H_0  \right].
\end{eqnarray}
The average charging power is given by the ratio of energy deposited on the battery during the charging process
and the time required to perform the unitary operation
\begin{eqnarray}
\label{power}
P(\tau) = \frac{W(\tau)}{\tau}.
\end{eqnarray}
This gives us a clear understanding of how to deposit energy as quickly as possible for this many-body system. The objective
of preparing such a battery is to maximize the extractable power, and hence it is important to choose a proper time interval
where the work deposited and average power reaches its maximum.

\section{Results and discussion}

In this section we investigate the charging performance of the quantum battery under different physical parameter regime.
We compute the work deposition $W(\tau)$ and the average power $P(\tau)$ using Eqs.~\eqref{work} and \eqref{power}
respectively. First we vary the interaction strength $g$ between the spins, after which we examine the battery performance
under different strengths of Landau-Zener driving field $v$. Finally, we explore the importance of anisotropy $\gamma$ in
the spin-chain and its role in efficient charging of the battery. We consider both nearest-neighbour \eqref{NN} and long-range
\eqref{LR} pairwise interactions of the Heisenberg XY spin-chain model \eqref{HB}.

\subsection{Charging performance of the quantum battery under different interaction strength}

\begin{figure*}[htp]
\centering
\begin{subfigure}[b]{0.45\textwidth}
\centering
\includegraphics[width=\textwidth]{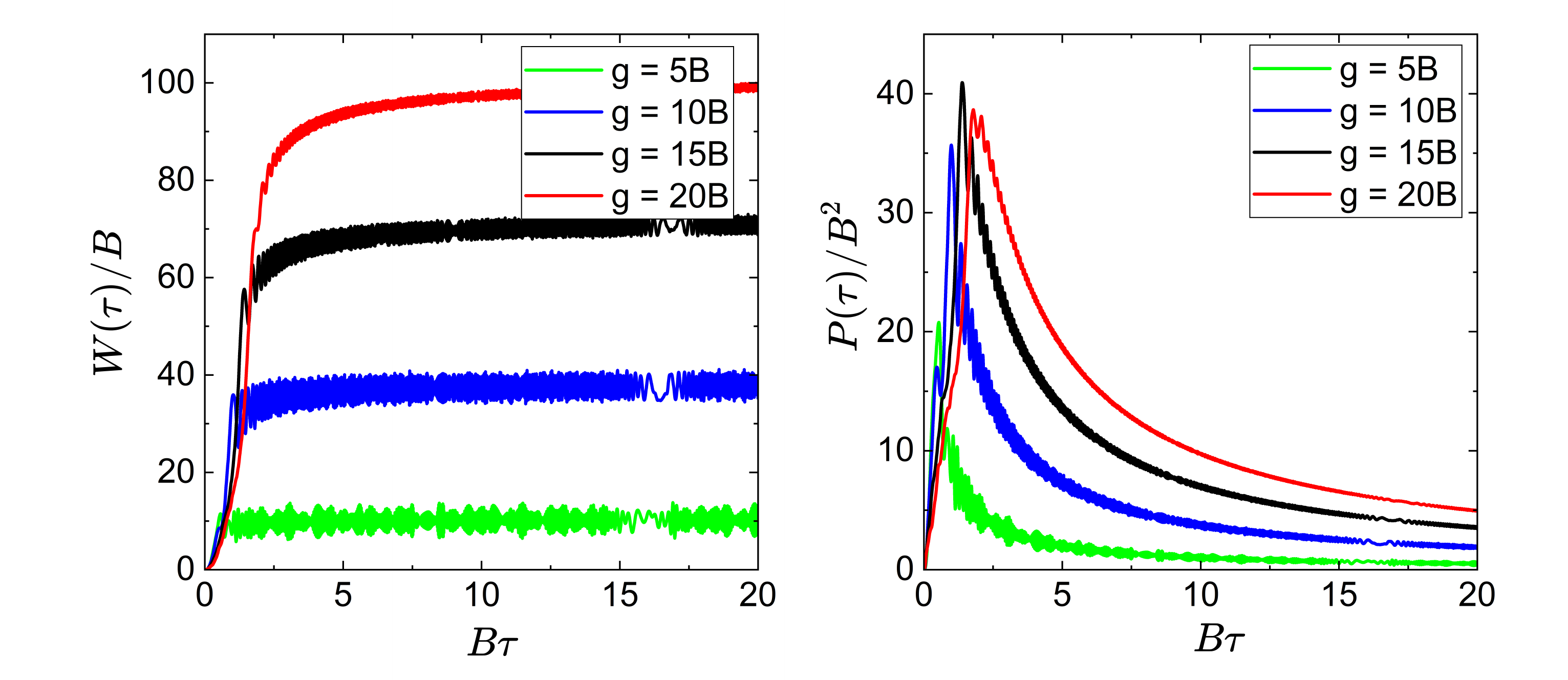}
\caption{Nearest-Neighbour interaction}
\label{fig1a}
\end{subfigure}
\hskip -0.0cm
\begin{subfigure}[b]{0.45\textwidth}
\centering
\includegraphics[width=\textwidth]{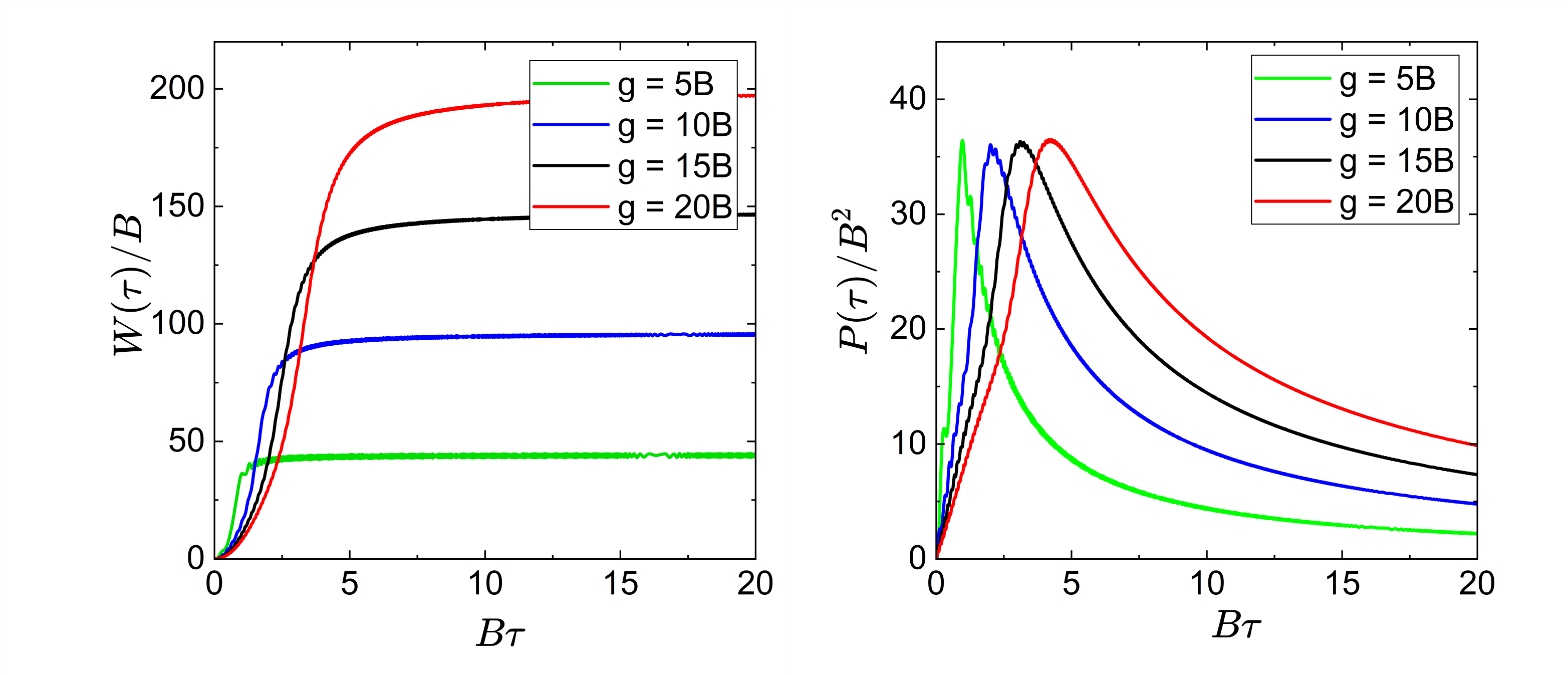}
\caption{Long-Range interaction}
\label{fig1b}
\end{subfigure}
\caption{We plot the work deposition $W(\tau)/B$ and average power $P(\tau)/B^2$ deposited for (a) nearest-neighbour and (b) long-range interaction for various values of interaction strength. Different curves (green for $g = 5B$, blue for $g = 10B$, black for $g = 15B$ and red for $g = 20B$) represent different values of $g$. The Landau-Zener driving field strength is taken as $v=10B$ for $N = 8$ spins and the magnetic field $B = 1$.}
\label{fig1}
\end{figure*}

In Fig.~\ref{fig1a} we plot the work deposition $W(\tau)$ and the average power $P(\tau)$ as a function of
charging time $\tau$, in a situation when the XY spins are interacting with their nearest-neighbours as described by
Eq.~(\ref{NN}). The dynamics of work deposition and the average power is examined for different interaction strengths.
Different curves in Fig.~\ref{fig1a} represent different coupling strengths between neighbouring spins, namely $g=5B$
(green line), $g=10B$ (blue line), $g=15B$ (black line), and $g=20B$ (red line). We fixed the Landau-Zener driving
strength at $v=10B$ and the anisotropy parameter is taken as $\gamma=0.5$. We observe that both the work
deposition $W(\tau)$ and the average power $P(\tau)$ increases very fast and approaches a maximum value in a very
short time scale. The advantage of using Landau-Zener driving is that the charging time needed to reach the maximum
$W(\tau)$ (or maximum average power $P(\tau)$) is very short. The battery is charged to its full capacity at $B\tau \sim 2$.
The maximum attainable value of $W(\tau)$ increases as we increase the coupling strength $g$. The average power
$P(\tau)$ is sustained for a longer time for higher values of $g$. Next, we demonstrate the charging capacity of the battery
for long-range interactions (\ref{LR}) in Fig. \ref{fig1b}. The qualitative behavior of $W(\tau)$ and $P(\tau)$ in this case
is somewhat similar to the charging dynamics of the battery with nearest-neighbour interaction. Under this long-range
interaction, the maximum value of $W(\tau)$ is also seen to enhance with increasing interaction strength $g$. Compared
to nearest-neighbor interactions, we observe a quantitative difference in the maximum value of $W(\tau)$ when considering
long-range interactions between the spins. However, the maximum value of the average power $P(\tau)$ remains nearly
unchanged while shifting from nearest-neighbor to long-range interaction. Interestingly, the work deposition $W(\tau)$
under weak interaction strengths reaches its maximum more quickly than under strong interactions (see the short-time
dynamics of $W(\tau)$ in Fig. \ref{fig1b} for different coupling strengths $g$), although that maximum value of $W(\tau)$
is lower when we consider weaker coupling strength.

In the nearest-neighbor configuration (Fig.~\ref{fig1a}), the interaction is localized: a given spin interacts only with its immediate neighbors. This will cause stronger quantum coherence effects where local quantum phases of individual spins are important features of the dynamics of the system. Such local interactions can exhibit quantum interference phenomena, where coherence between different parts of the system leads to constructive or destructive interference in energy exchange \cite{bose2003quantum}. In the long time evolution, the finite correlation range limits the information spread throughout the system, which may result in revivals or fluctuations of energy deposition, explaining the oscillations that occur in the average power and work deposition in the NN model. On the contrary in Fig. \ref{fig1b}, the long-range interaction model allows many spins to interact simultaneously resulting to average out local fluctuations, local coherence effects at individual sites are reduced. Such a scenario would cause quantum information to propagate in an extremely smooth manner in the spin chain, thereby spreading energy almost uniformly along the spin chain and depositing work and power in an efficient way. This long-range complex interaction suppress local quantum interference effects and leads to a more uniform dynamics; that's why we don't see fluctuations of the same nature in the long-range model \cite{casanova2012quantum}.

\begin{figure*}[htp]
		\centering
		\begin{subfigure}[b]{0.45\textwidth}
			\centering
			\includegraphics[width=\textwidth]{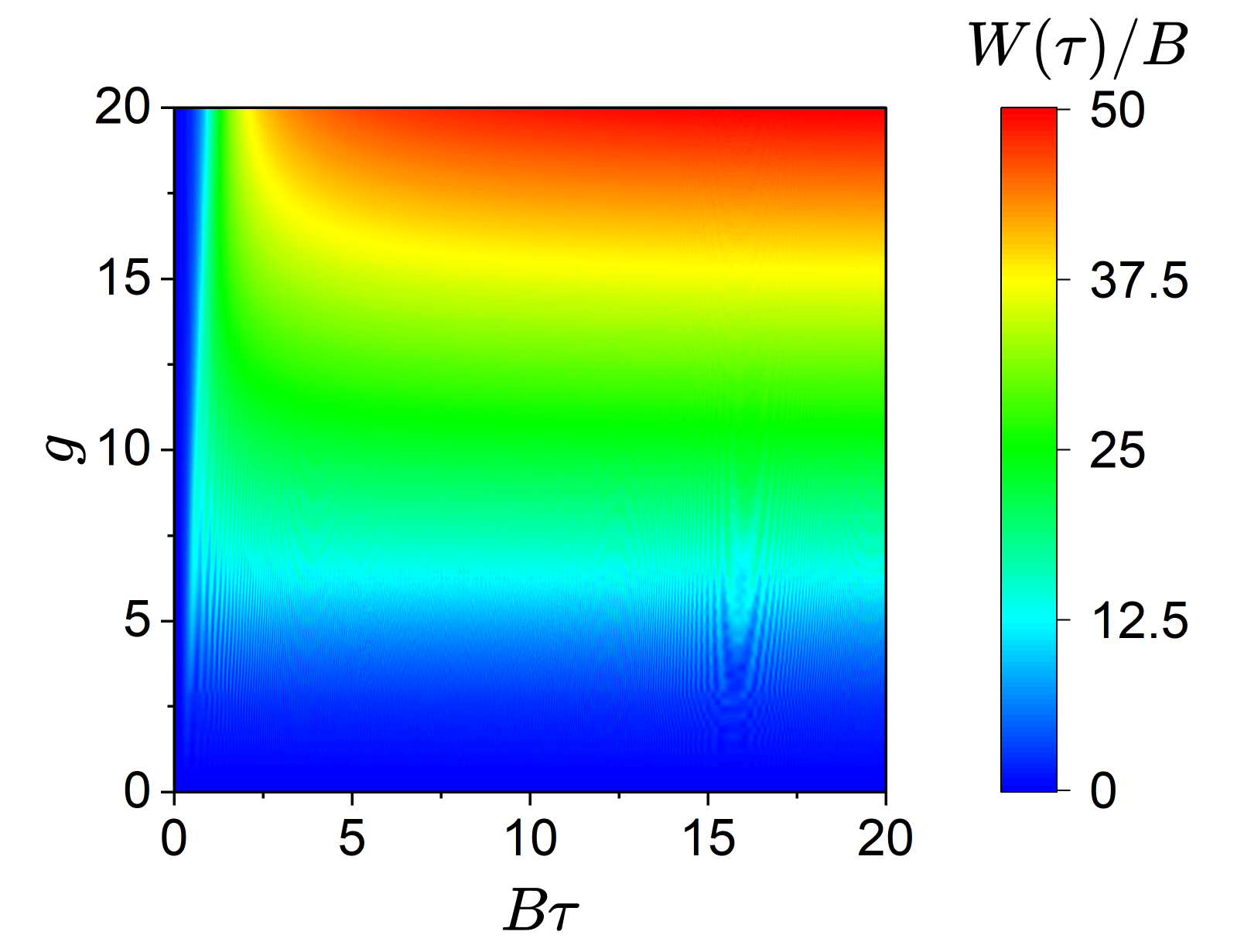}
			\caption{Work deposited in Nearest - Neighbour}
			\label{fig2a}
			\end{subfigure}
		\hskip -0.0cm
		\begin{subfigure}[b]{0.45\textwidth}
			\centering
				\includegraphics[width=\textwidth]{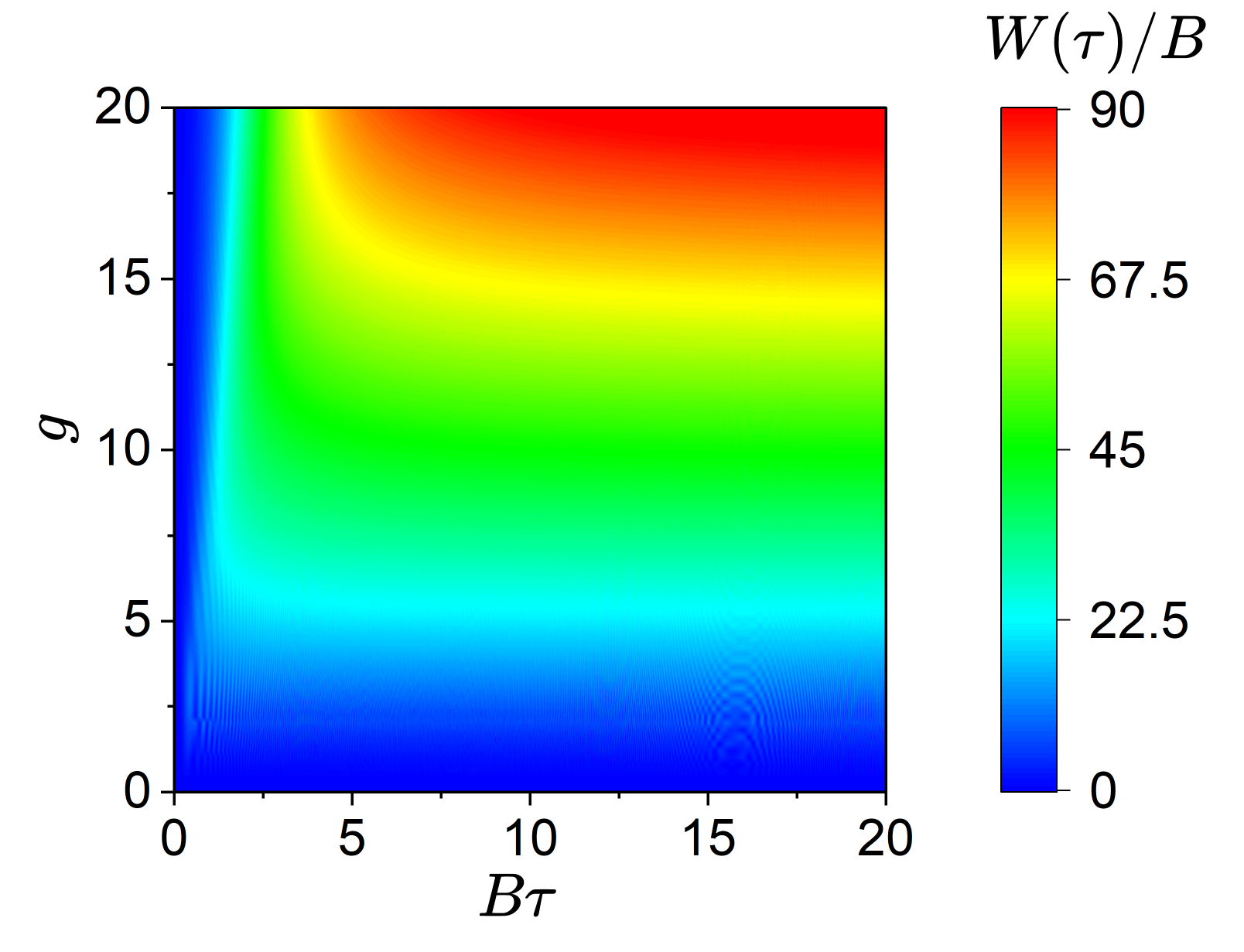}
			\caption{Work deposited in Long- Range}
			\label{fig2b}
		\end{subfigure}
		\caption{The contour plot of work deposition as a function of charging time $\tau$ and coupling strength $g$ for (a) nearest-neighbour
and (b) long-range interaction. We have considered a range of coupling strength $g$ from $0$ to $20B$. The Landau-Zener driving field
strength is taken as $v=10B$ for $N=7$ spins.}
		\label{fig2}
	\end{figure*}

	To present a comprehensive phase diagram (see Fig.~\ref{fig2}), we illustrate the dynamics of work deposition as a function of charging time $B\tau$ and coupling strength $g$ for both nearest-neighbour (\ref{NN}) and long-range interactions (\ref{LR}). We consider a continuous range of interaction strengths $g$ from $0 \leq g \leq 20B$. In Fig.~\ref{fig2a}, we plot the work deposition for the nearest-neighbour configuration. We observe that the battery starts to charge after a very short time interval ($B\tau \sim 1$), and as we increase the coupling strength between the nearest-neighbour spins, the charging rate rapidly enhances. Similarly, in Fig.~\ref{fig2b}, we see analogous work deposition characteristics for long-range interactions. However, the magnitude of work deposition is significantly higher in this case than in the nearest-neighbour interaction. In both cases (Figs.~\ref{fig2a} and \ref{fig2b}), for a given charging time, the value of $W(\tau)$ shows three distinct regions in the color map. In the spectrum of interaction strength, the first region is blue from $g=0$ to $g \sim 5B$, followed by the green region from $g \sim 5B$ to $g \sim 15B$, and the red region from $g \sim 15B$ to $g \sim 20B$. The first transition level in energy deposition occurs at $g \sim 5B$, and the second transition level occurs at $g \sim 15B$. We observe that the charging efficiency improves as the interaction strength among the spins increases in both the long-range and nearest-neighbour cases, as evidenced in Fig.~\ref{fig2}.

\subsection{Detailed Exploration of Landau-Zener Driving Field in Many-body Spin Chain Battery}
	
	Landau-Zener driving in quantum mechanics is used to describe the transition that occurs between two quantum states, basically as a result of changing energy levels due to a linearly varying external driving field \cite{zener1932non,landau1932theorie}. However, it applies specifically to a system that has two energy levels, or what is called a two-level system. In this context, the two levels are by no means "closely separated" in energy at all times; rather, the energy gap may vary, often crossing over an avoided crossing, where the transition probability between the two states is nonzero. The traditional model supposes that the time-dependent Hamiltonian is chosen so that the difference between the two energies changes linearly in time \cite{wang2008landau}. We investigate the role of this type of Landau-Zener driving in our spin system.

\begin{figure*}[htp]
		\centering
		\begin{subfigure}[b]{0.45\textwidth}
			\centering
			\includegraphics[width=\textwidth]{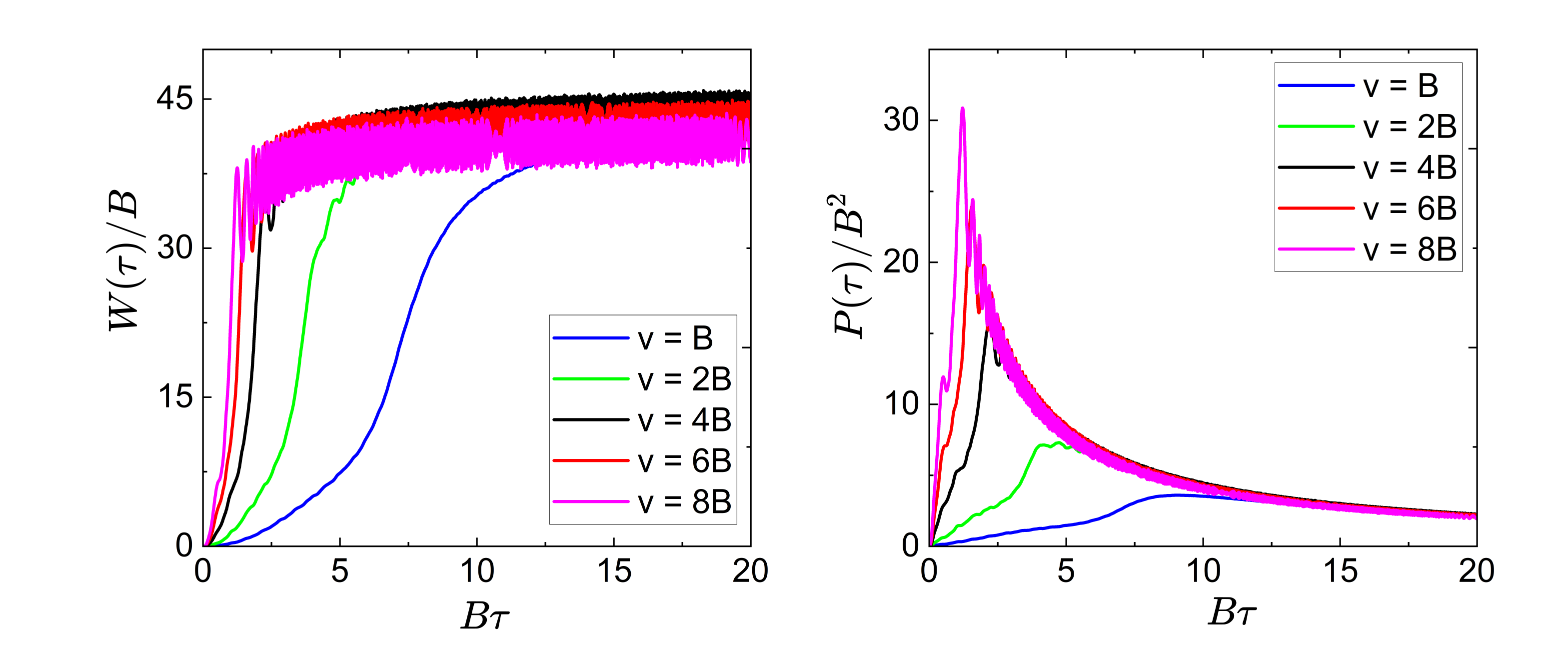}
			\caption{Nearest-Neighbour interaction}
			\label{fig3a}
		\end{subfigure}
		\hskip -0.0cm
		\begin{subfigure}[b]{0.45\textwidth}
			\centering
			\includegraphics[width=\textwidth]{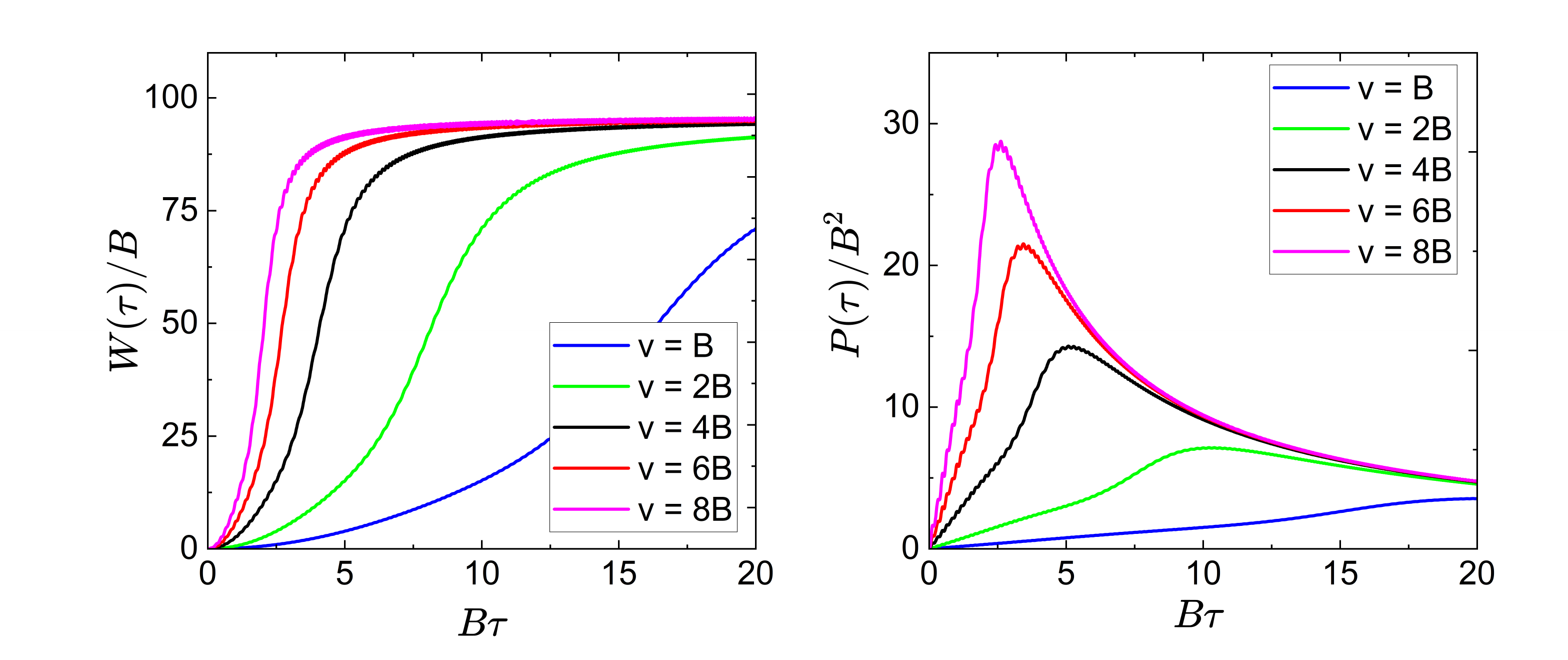}
			\caption{Long-Range interaction}
			\label{fig3b}
		\end{subfigure}
		\caption{Exact numerical plots for the work deposition $W(\tau)/B$ and average power $P(\tau)/B^2$ deposited for the nearest-neighbour and long-range interaction for various values of Landau-Zener driving , $v$ (magenta line $v = 8B$, red line $v = 6B$, black line $v = 4B$, green line $v = 2B$ and blue line $v = B$) for the magnetic field $B = 1$ and the coupling strength $g = 10B$ and the anisotropy parameter $\gamma = 0.5$ for $N = 8$ spins.}
		\label{fig3}
	\end{figure*}

	In Fig.~\ref{fig3a}, we demonstrate the influence of varying the Landau-Zener driving amplitude $v$ on the dynamics of work deposition and average power for the nearest-neighbour interaction model of the XY spin chain. We plot the work deposition $W(\tau)$ and average power $P(\tau)$ against charging time $\tau$ for five different driving strengths $v$. The curves in Fig.~\ref{fig3a} represent various field strengths of the Landau-Zener driving: $v=B$ (blue line), $v=2B$ (green line), $v=4B$ (black line), $v=6B$ (red line), and $v=8B$ (magenta line). The many-body coupling strength is set to $g=10B$, and the anisotropy parameter is $\gamma=0.5$. From Fig.~\ref{fig3a}, we observe that work deposition becomes highly fluctuating over time, especially at higher $v$ values. These fluctuations (notably for $v=4B$, $v=6B$, $v=8B$) may indicate non-adiabatic transitions or energy redistribution due to nearest-neighbour coupling. This reflects a more complex and less smooth energy exchange mechanism compared to the long-range interaction case. Using the same parameter values for $g$ and $\gamma$, we examine work deposition and average power dynamics in Fig.~\ref{fig3b} for long-range interaction. We note the smooth rising of work deposition curves for all values of $v$ (Fig.~\ref{fig3b}), which saturate in the end. It takes longer time to reach the saturation value when we consider lower value of driving strength (see the blue line in Fig.~\ref{fig3b} for $v=B$). For higher values of $v$ (black line is for $v=4B, $ red is for $v=6B$, and magenta is for $v=8B$), the work deposition is faster and saturation takes place at an earlier time, although the qualitative behaviour of the charging dynamics remains similar as that with nearest-neighbour interaction. The saturation occurring at longer time tells us that the many-body system is excited to some state defined by the steady-state limit. It is related to the adiabatic limit where the system cannot accept more energy because the system has already reached to its maximum energy configuration. For all $v$, the average power $P(\tau)$ has a sharp maximum that decreases and gradually stabilizes. With higher values of $v$ (see Fig.~\ref{fig3b} for $v=8B$), the maximum power output is significantly increased. A stronger driving field induces faster initial energy deposition, leading to greater power output in a shorter time. The charging time should be optimally selected because of the decaying nature of power, the battery should be charged until the maximum work is deposited within the system. Analyzing the results in Fig.~\ref{fig3}, we notice that for nearest-neighbour case (Fig.~\ref{fig3a}) the fluctuation is prominent and saturation still occurs for higher values of $v$, but the maximum work deposition is lower than that of the long-range interaction model (see Fig.~\ref{fig3b}). The power output ($P(\tau)$) also shows sharp peaks for higher $v$ values. The decay in power is more abrupt in the nearest-neighbor model, indicating that the system rapidly loses its ability to sustain power output after the initial surge, likely due to the localized interactions limiting energy flow across the system. For this specific charging protocol, we also see (Figs.~\ref{fig3a} and \ref{fig3b}) that by increasing $v$ consistently leads to a faster work deposition and higher power peaks in both models, but the nearest-neighbor model is more sensitive to driving amplitude in terms of energy fluctuations. The long-range model
	handles increasing $v$ more effectively, maintaining smoother work deposition and more stable power output.

\begin{figure*}[htp]
		\centering
		\begin{subfigure}[b]{0.45\textwidth}
			\centering
			\includegraphics[width=\textwidth]{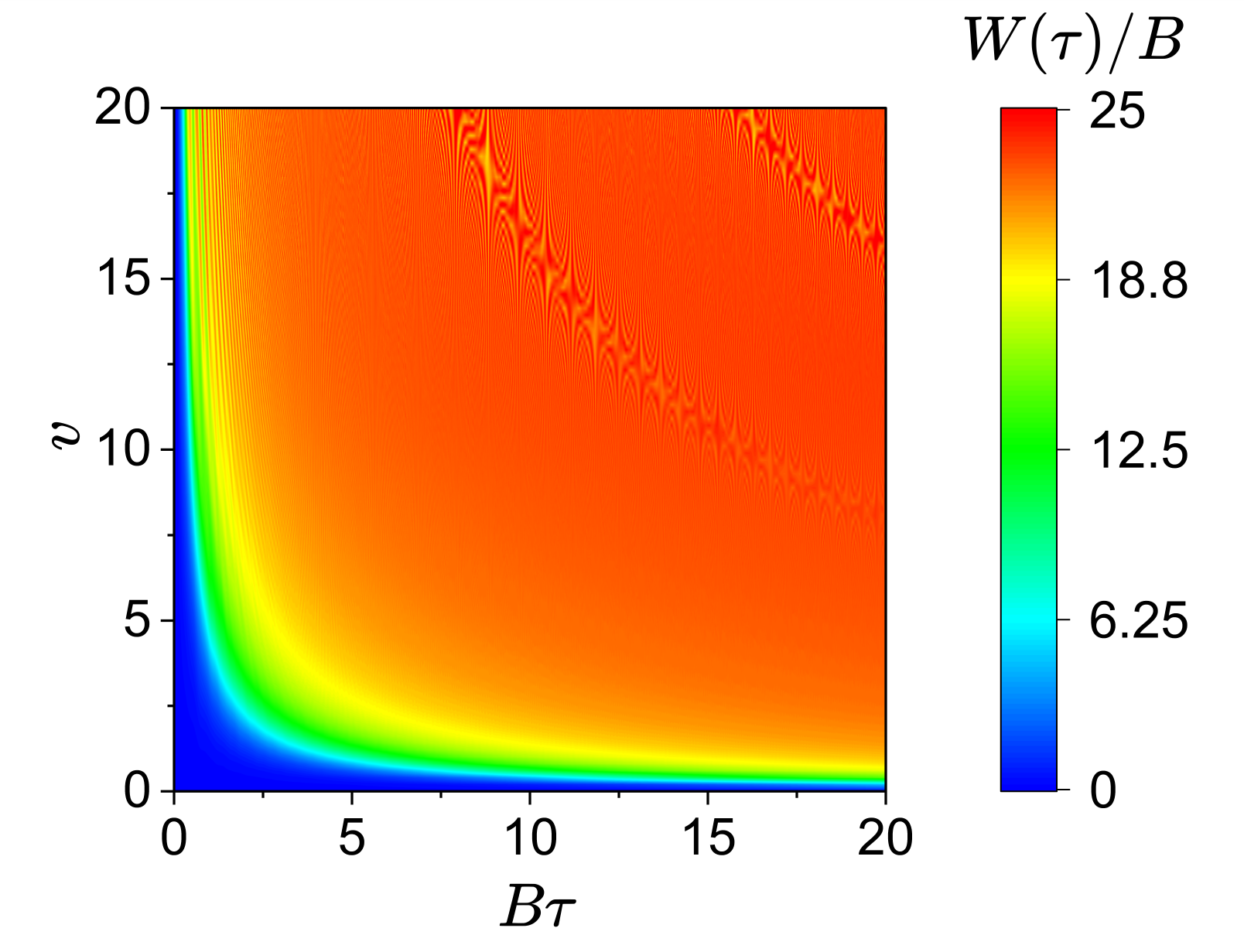}
			\caption{Work deposition in Nearest - Neighbour}
			\label{fig4a}
		\end{subfigure}
		\hskip -0.0cm
		\begin{subfigure}[b]{0.45\textwidth}
			\centering
			\includegraphics[width=\textwidth]{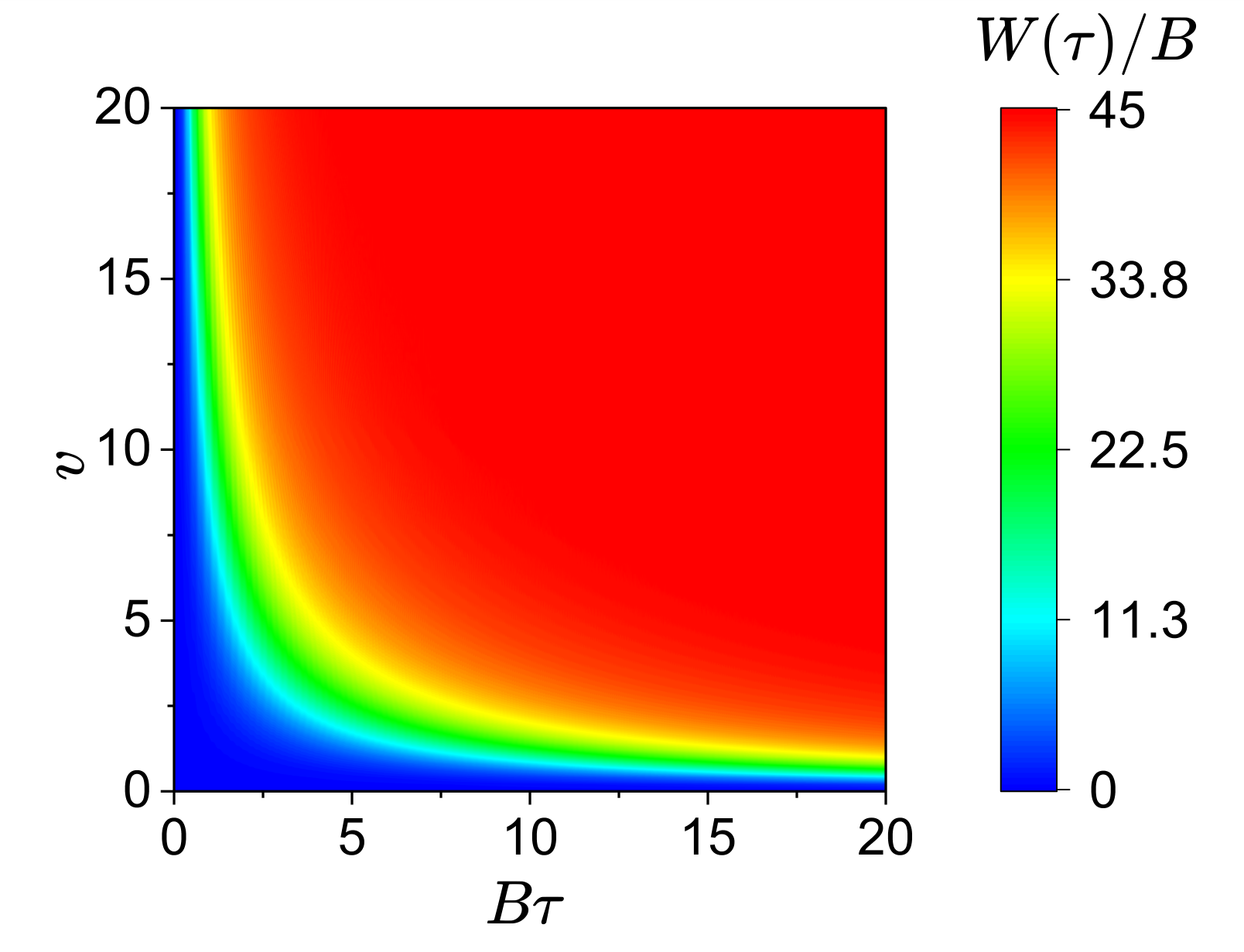}
			\caption{Work deposition in Long- Range}
			\label{fig4b}
		\end{subfigure}
		\caption{The work deposition contour plot for (a) Nearest-neighbor and (b) Long-range interactions as a function of charging time $B\tau$ and driving strength $v$. A range of driving strength $v$ from $0$ to $20B$ has been taken into consideration. The coupling strength is fixed at $g = 10B$, the anisotropy parameter $\gamma = 0.5$ with the number of spins, $N = 7$.}
		\label{fig4}
	\end{figure*}

	To provide a more comprehensive visual illustration (refer to Fig.~\ref{fig4}), we detail the dynamics of energy deposition relative to charging duration $B\tau$ and the LZ-driving parameter $v$ for both nearest-neighbour (\ref{NN}) and long-range interactions (\ref{LR}). We investigate a continuous range of driving parameter strengths $v$ from $0 \leq v \leq 20B$. In Fig.~\ref{fig4a}, we present the energy deposition for the nearest-neighbour configuration. We observe that the battery starts to charge after a brief interval ($B\tau \sim 1$), and as the driving parameter $v$ from the external driving increases, the charging rate accelerates significantly. From the plot in Fig.~\ref{fig4a}, we see that the increase of the work deposition with $v$ and $B\tau$ is quite smooth. It is also clear with a gradient of color going from blue to red, indicating lower values at the bottom left to higher values at the top right. Here, in Fig.~\ref{fig4a} the work deposition $W(\tau)$ is clearly $v$ and $B\tau$ dependent, but the structure is distinctly different. This plot has clearly visible oscillations, or an interference-like pattern in the top-right corner of the panel. These oscillations tend to reflect a more complicated form of dynamics in work deposition for nearest-neighbor interactions which can be partly related to energy transfer effects or quantum coherence phenomena that appear much more pronounced if only nearest neighbors interact. The oscillations in the first graph reflect interference and dynamic effects, where energy deposition becomes less uniform, thus possibly finite-size effects and coherence. This is also due to nearest-neighbour interaction that makes the energy exchange local, which then is responsible for the observed complex pattern of the work deposition. Similarly, in Fig.~\ref{fig4b}, we see analogous work deposition characteristics for long-range interactions. The extent of work deposition is notably higher in this context compared to the nearest-neighbour interaction. The smooth flow indicates that the work deposition is steady with both parameters; this means that the energy deposition for the long-range interaction model is more or less uniform. Long-range interaction (Fig.~\ref{fig4b}) yields much higher overall work deposition, going up to 45, whereas the nearest-neighbor interaction (Fig.~\ref{fig4a}) peaks at a much lower value of 25. Therefore, it results in the indication that long-range interactions are more effective for depositing energy into the system.  The fact that the second plot (see Fig.~\ref{fig2b}) is oscillation free and the gradient is smooth implies that long-range interactions are more effective to an extent in distributing energy uniformly across the chain and hence a stable increase in work.

	\subsection{The Role of Anisotropy in Spin-Chains: Implications for Quantum Battery Efficiency}
	
\begin{figure*}[htp]
		\centering
		\begin{subfigure}[b]{0.45\textwidth}
			\centering
			\includegraphics[width=\textwidth]{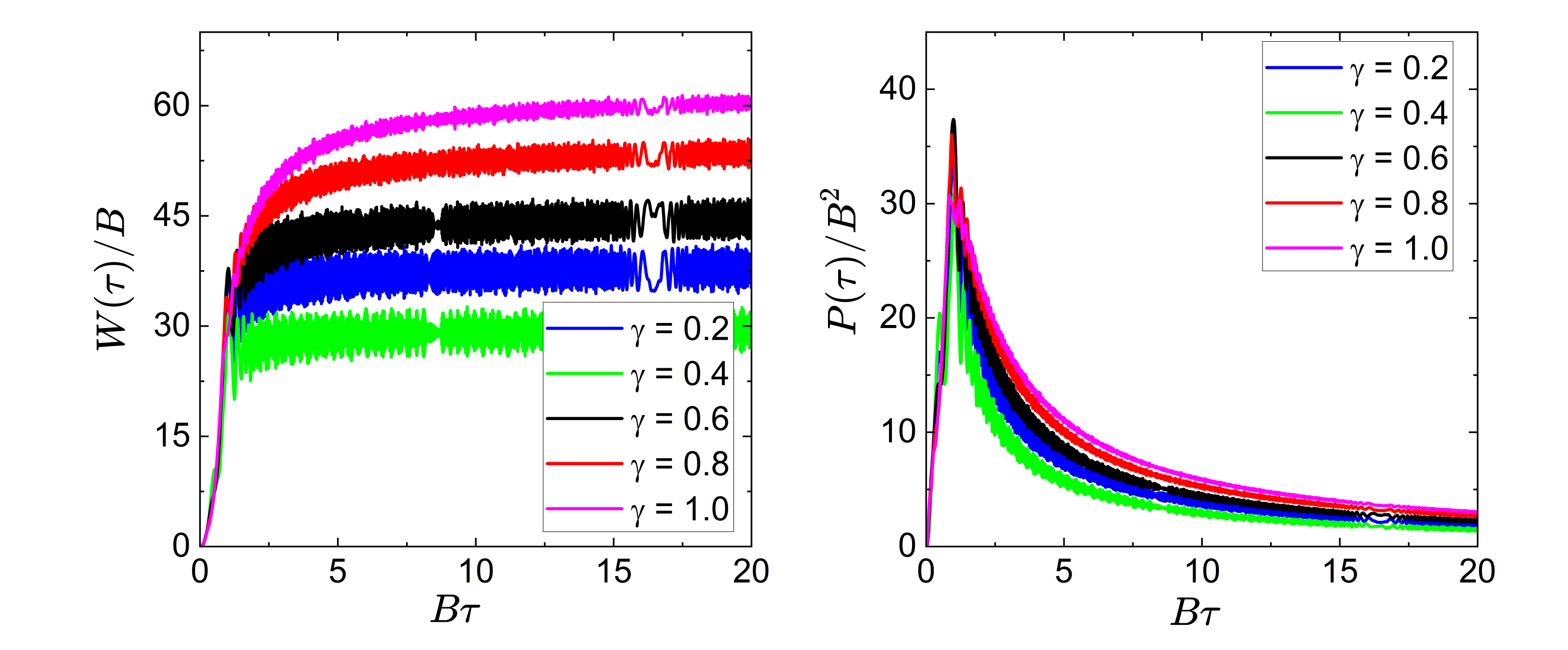}
			\caption{Nearest-Neighbour interaction}
			\label{fig5a}
		\end{subfigure}
		\hskip -0.0cm
		\begin{subfigure}[b]{0.45\textwidth}
			\centering
			\includegraphics[width=\textwidth]{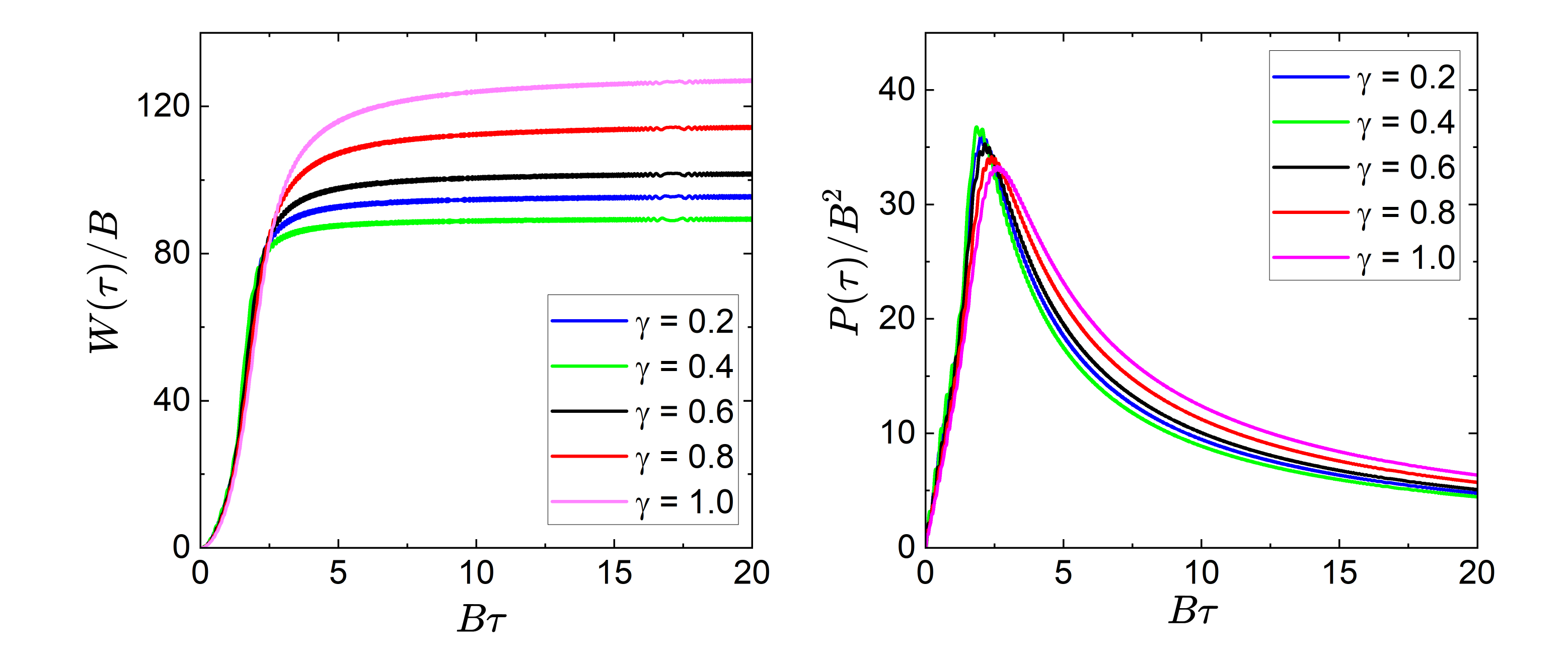}
			\caption{Long-Range interaction}
			\label{fig5b}
		\end{subfigure}
		\caption{We have plotted the work deposition $W(\tau)/B$ and average power $P(\tau)/B^2$ deposited for the nearest-neighbour and long-range interaction for different values of anisotropy (magenta line $\gamma = 1.0$, red line $\gamma = 0.8$, black line $\gamma = 0.6$, green line $\gamma = 0.4$ and blue line $\gamma = 0.2$) for the magnetic field $B = 1$ and the coupling strength $g = 10B$ and the driving field strength $v = 10B$ for $N = 8$ spins.}
		\label{fig5}
	\end{figure*}

	In this section, we try to analyze how much of importance anisotropy is in the context of spin chains and how does it affects in the work deposition of the system. From Fig.~\ref{fig5}, we illustrate how changes in anisotropy affect the work deposition dynamics and average power within the nearest-neighbour and long-range interaction model of the XY spin chain. We have calculated numerically and plotted the values of work deposition $W(\tau)$ and average power $P(\tau)$ against charging time $B\tau$ for five distinct anisotropy values $\gamma$. The curves in Fig.~\ref{fig5} depict various anisotropy levels from the XY Hamiltonian (\ref{H0}): $\gamma=0.2$ (blue line), $\gamma=0.4$ (green line), $\gamma=0.6$ (black line), $\gamma=0.8$ (red line), and $\gamma=1.0$ (magenta line). The many-body coupling strength is fixed at $g=10B$, and the Landau-Zener driving parameter is set to $v=10B$.

	In Fig.~\ref{fig5a} for the nearest neighbour interaction, as $\gamma$ is increased (from $0.2$ to $1.0$), the final work deposition increases, as can be seen from the coloured lines moving up. The magnitude of work deposition is lower for the nearest-neighbor case from Fig.~\ref{fig5a}. Interestingly, the oscillations are much stronger and less periodic for the nearest-neighbor case, especially at larger $\gamma$, which might suggest that the nearest-neighbor interaction is less stable about storing energy. Actually, for larger $\gamma$ values, the erratic behavior continues with oscillations of small amplitude for a longer time. The peak power (see the second plot in Fig.~\ref{fig5a}) is a little higher for $2 < B\tau <3$. A more interesting observation is that the work deposition for $\gamma = 0.2$ (blue line) is higher than that for $\gamma = 0.4$ (green line) while the other values of work deposition increases with increase in $\gamma$. Long-range interaction (see Fig.~\ref{fig5b}) obviously allows for a greater total amount of work deposition, so that the quantum battery can store energy corresponding to a greater amount using the same external driving field. For larger $\gamma$, the system deposits more work and achieves higher values, but saturation behaviour is consistent in different $\gamma$ values after $B\tau \approx 10$. The oscillations around $5 < B\tau <10$ are more pronounced for a larger $\gamma$, which suggests a stronger energy exchange and quantum coherence in the system. The maximum power deposition (look for the second plot in Fig.~\ref{fig5b}) occurs around $2 < B\tau < 3$, after which it decreases sharply. Higher $\gamma$ values yield higher peak power deposition, indicating that stronger anisotropy enhances the charging dynamics initially. However, after $B\tau \approx 5$, the power decreases and remains close to zero for all values of $\gamma$, indicating that the energy stored remains stable over time, with minimal further power input. An intriguing similar observation in LR interaction also is that the work deposition for $\gamma = 0.2$ (blue line) is higher than for $\gamma = 0.4$ (green line) whereas for other values the work deposition increases with a rise in $\gamma$.

	For instance, $W(\tau)/B$ for $\gamma = 1.0$ saturates around the value $60$ (Fig.~\ref{fig5a}), whereas in the long-range case(Fig.~\ref{fig5b}), it saturates around $120$.  On the other hand, the nearest-neighbor interaction (see Fig.~\ref{fig5a}) appears to limit the system's storage capacity. This restriction is due to the limited correlations between the spins, resulting in a less efficient charging process. The power fall-off from the peak (see the second plot in Fig.~\ref{fig5a}) is more rapid compared to the long-range interaction (see the second plot in Fig.~\ref{fig5b}). The behavior of both models is quite similar near the peak power deposited as a function of time. Long-range interaction clearly shows a smoother decay compared to the high fluctuation as seen in the behavior of the nearest-neighbor interaction, which suggests the dominance of fluctuations in the energy transfer has more to do with reduced range in spin-spin coupling within the system of nearest neighbors. As $\gamma$ increases, work deposition and average power also increase, but the latter grows stronger in the long-range interaction model. The oscillatory behaviour of charging and average power is prominent in the nearest-neighbour interaction case. From Fig.~\ref{fig5} we see an enhanced work deposition and average power resulting from a richer dynamics of energy exchange.

\begin{figure*}[htp]
		\centering
		\includegraphics[width=\textwidth, height=0.5\textheight]{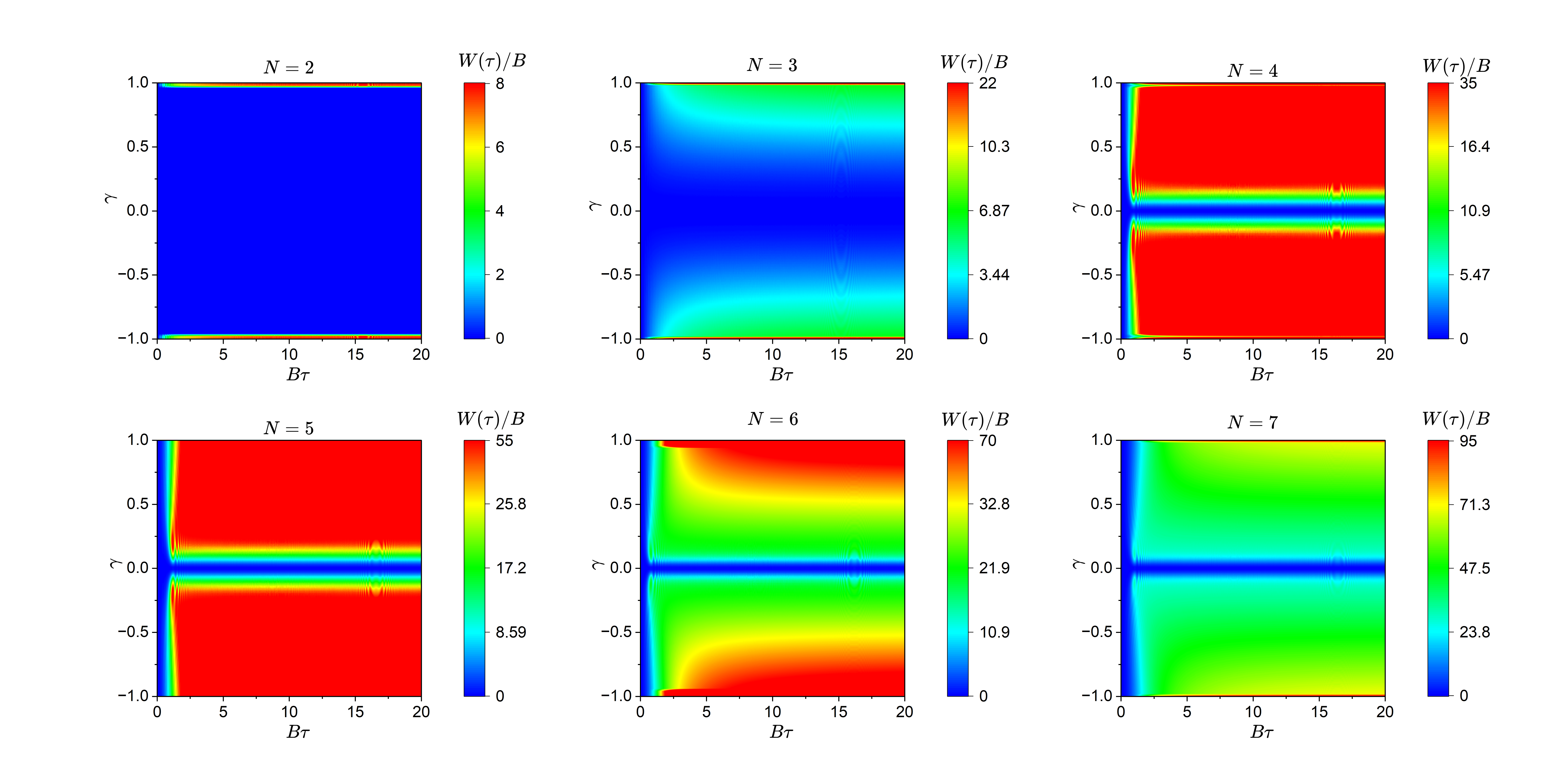}
		\caption{The contour plot of work deposition $W(\tau)/B$ as a function of charging time $B\tau$ and the anisotropy parameter $\gamma$ in the long-range interaction with varying number of spins. Anisotropy parameter $\gamma$ is ranging from $-1$ to $+1$. The coupling strength is $g = 10B$ and the Landau-Zener driving field strength is $v=10B$.}
		\label{fig6}
	\end{figure*}

		We analyze the kinetics of energy deposition concerning the charging time $B\tau$ and the anisotropy parameter $\gamma$ within the range $-1 \leq \gamma \leq +1$ for long-range interactions (\ref{LR}), providing a more comprehensive visual representation (see Fig.~\ref{fig6}). We consider investigating only for the long-range interaction in this exploration. The motivation for investigating long-range interactions stems from our previous results, which demonstrated that long-range interactions exhibit greater efficiency compared to nearest neighbor interactions. This conclusion is based on a comparative analysis of both interaction types, highlighting the superior performance of long-range interactions. To understand how the `charging' behavior changes as the number of spins $N$ increases, we aim to investigate a continuous range of anisotropy $\gamma$ from	$-1 \leq \gamma \leq +1$ for various values of $N$. The results from Fig.~\ref{fig6} explores increasing values of $N$ and how it impacts in charging process. When there are only 2 spins, the system shows a uniform work deposition with minimal response to changes in	$B\tau$ and $\gamma$. $W(\tau)$ remains close to zero for all values of $\gamma$ except $\gamma = \pm 1$. This is likely due to the simplicity of a two-particle system, where there is limited room for interaction. When we increase to 3 spins a clear dependence on $\gamma$ starts to appear. For $\gamma \approx \pm 1$, the system shows slightly more work deposition as compared to $\gamma = 0$, but the values of $W(\tau)/B$ remain relatively low ($\sim$ 3 to 7). This suggests that with three particles, the anisotropy has a mild effect, but there is still limited dynamics in this small system. For $N = 4$ spins, The work deposition pattern changes significantly. Around $\gamma =0$ we see a sudden jump of work deposition as we increase the charging time $B\tau$. A distinct central band forms around $\gamma = 0$, suggesting that in this region, work deposition is lower compared to the case of higher values of $\gamma$. The increased system size introduces more complex behavior with stronger dependence on $\gamma$ and $B\tau$. While looking for $N = 5$ spins, similar behavior to $N = 4$, but the magnitude of work deposition overall increases, reaching higher values ($\sim 55$ for the maximum).	The central band near $\gamma = 0$ becomes more prominent, with very low deposition around zero and strong deposition around $\gamma = \pm 1$. This indicates that as the system size increases, the anisotropy starts playing a major role in the work deposition. While this continues, we explore for $N = 6$ and we observe that the work deposition increases further, with values reaching up to $\sim 70$. As the trend continues: $W(\tau)$ is lowest around $\gamma = 0$ and significantly higher as we increase the anisotropy $\gamma$. The central band is well-defined, with work deposition more evenly distributed at higher $B\tau$ values. Finally at $N = 7$ we see that the work deposition continues to increase, reaching values up to $\sim 95$. The central region near $\gamma = 0$ continues to show low deposition, but the boundaries near $\gamma = \pm 1$ exhibit the strongest deposition. The system shows increasingly complex behavior with more pronounced effects of anisotropy as $N$ increases.

		As the number of particles $N$ increases, the work deposition increases significantly. This is expected as larger systems allow for more interactions and more complex dynamics. For larger $N$, the anisotropy parameter $\gamma$ starts to play a crucial role. Work deposition is highest near $\gamma = \pm 1$ and lowest near $\gamma = 0$. This suggests that the system’s response to external fields becomes more sensitive to anisotropy as the number of particles increases. For smaller values of $N$, the system is less responsive to variations in $B\tau$. However, as $N$ increases, the work deposition increases with $B\tau$, and the system develops more complex behavior in response to time evolution. The system shows minimal work deposition in this limit $(\gamma = 0)$, as isotropy reduces the effectiveness of the external drive on spin alignment. The system shows strong work deposition in these limits $(\gamma = \pm 1)$, suggesting that the external field is more effective in driving spin alignment in anisotropic regimes. This is likely because the anisotropic interactions lead to stronger spin correlations, which are more easily influenced by external fields.

	\section{Comparing a periodic driving with Landau-Zener driving}
	
	Recent studies have explored the implementation of periodic potentials as external charging fields for quantum batteries \cite{mondal2022periodically, guo2024analytically}. This innovative approach aims to enhance the efficiency and performance of quantum battery systems by leveraging the unique properties of such potentials . In this section, we take a sinusoidal form of a periodic potential and do a comparative study on the maximum work deposited, $W_{max}$ with the LZ (linear) driving. The objective is to ascertain whether the periodic potential or the linear driving enhances charging performance and to evaluate the impact of increasing the number of spins $N$ on this process. We apply a periodic time dependent external field replacing the linear driving (LZ) of the form,
	
	\begin{eqnarray}
		V(\tau) = v\sin(\omega \tau)\sum_{i=1}^{N} \sigma_{i}^{z}
	\end{eqnarray}
	where $v$ is our driving parameter for our periodic external driving field which is analogous to the Eqn. \eqref{LZ} except that now we have sinusoidal time-dependence with $\omega$ being frequency of the external field.
	
\subsection{Maximum work deposition in spin chain quantum batteries under strong coupling}

\begin{figure*}[htp]
		\centering
		\begin{subfigure}[b]{0.45\textwidth}
			\centering
			\includegraphics[width=\textwidth]{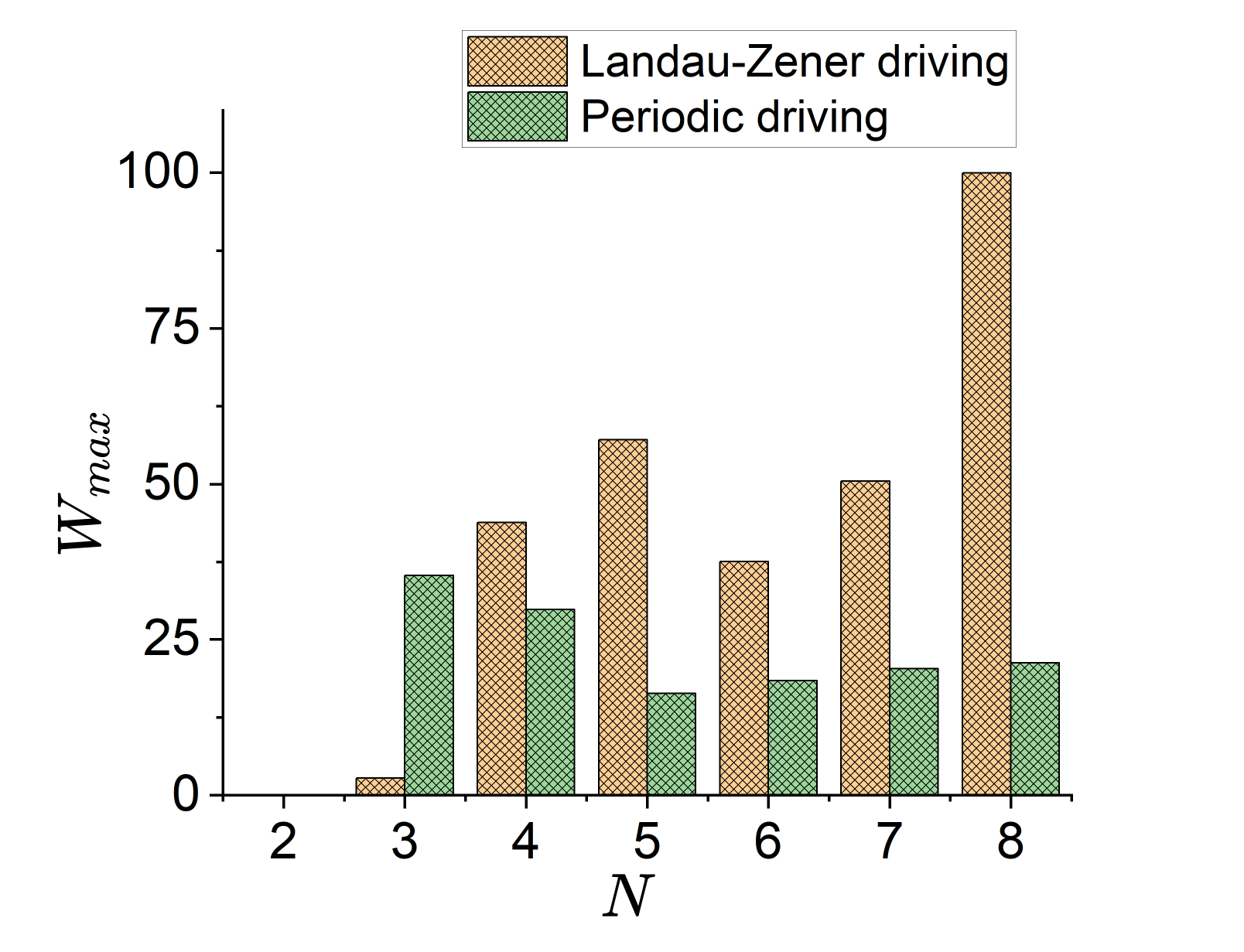}
			\caption{Nearest-Neighbour interaction}
			\label{fig7a}
		\end{subfigure}
		\hskip -0.0cm
		\begin{subfigure}[b]{0.45\textwidth}
			\centering
			\includegraphics[width=\textwidth]{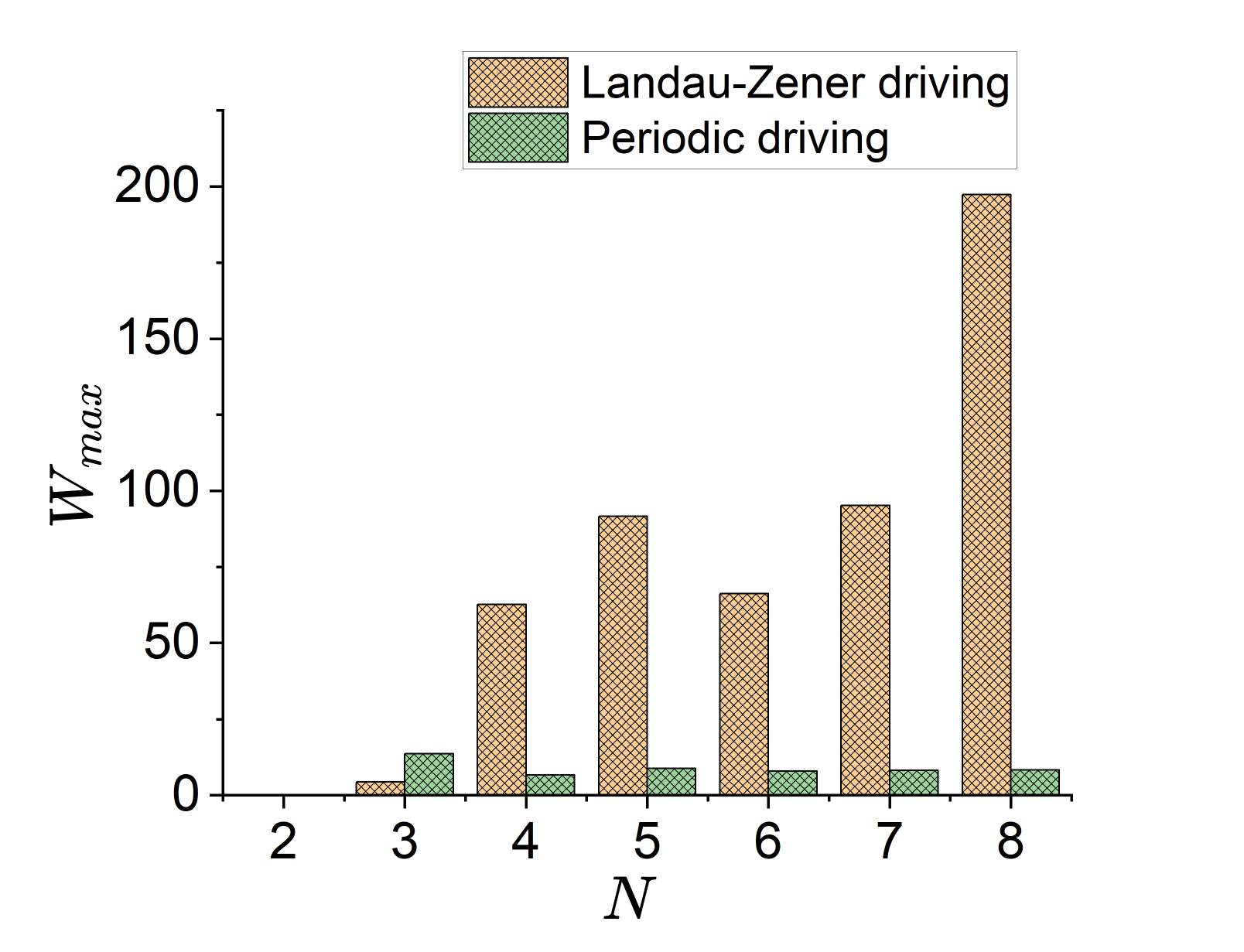}
			\caption{Long-Range interaction}
			\label{fig7b}
		\end{subfigure}
		\caption{We have taken the maximum value of $g$ and plot for the maximum value of $W$ in the over all time range $0 \leq \tau \leq 20$ for increasing values of the number of spins. $W_{max}$ Vs $N$ calculated for LR and NN. The parameters  are $g = 20B$, $\gamma = 0.5$, $\omega = 4B$ and $v = 10B$. The periodic driving and linear driving values are compared with respect to long-range (LR) and nearest-neighbour (NN) interaction.}
		\label{fig7}
	\end{figure*}

Here, we make a comparative analysis of charging with respect to periodic and Landau-Zener driving. We consider the maximum allowed value of the coupling strength to be $g=20B$. In Fig.~\ref{fig7}, we show how increasing the number of spins $N$ affects the maximum work deposited for the nearest-neighbor interaction (see Fig.~\ref{fig7a}) and the long-range interaction (see Fig.~\ref{fig7b}). We have taken the maximum value of $W$ in the charging time range $0\leq B\tau \leq 20$ for the XY spin chain. For various choices of $N$, we depict the maximum work deposited $W_{max}$ during this charging period $B\tau$. The bar diagrams in Fig.~\ref{fig7a} and Fig.~\ref{fig7b} depict the nearest-neighbour interaction and long-range interaction, respectively, for increasing values of $N$. These figures also highlight the maximum values of work deposition for both scenarios. We consider both linear and periodic driving. The linear driving (LZ-driving), is represented by the brown column bar, while periodic driving is represented by the green column bar. The many-body coupling strength is set to $g=20B$, the anisotropy parameter is $\gamma=0.5$, the amplitude of driving field $v=10B$ is taken for both Landau-Zener and periodic driving. The frequency of the periodic driving is set to $\omega = 4B$ for both NN (Fig.~\ref{fig7a}) and LR (Fig.~\ref{fig7b}) scenarios. The maximum work deposition $W_{max}$ typically rises with the number of spins, $N$ in both LR and NN cases, exhibiting a notable increase, particularly at $N = 8$. The increase in $W_{max}$ with $N$ in the NN interaction plot Fig.~\ref{fig7a} is less pronounced than in LR, indicating that the NN case's energy deposition is more restricted by the narrower interaction range. Across all values of $N$, linear driving consistently yields a higher $W_{max}$ than periodic driving (see Fig.~\ref{fig7}). At higher $N$ values, the difference becomes more pronounced. We now consider specifically the case of $N=8$ spins (see Fig.~\ref{fig7a}) for nearest neighbour interaction. In this case, the value of $W_{max} \sim 100$ for LZ-driving and $W_{max}\sim 20$ for periodic driving.
	
	 For the Landau-Zener (LZ) scenario (linear driving) in Fig.~\ref{fig7b} under LR interaction, $W_{max}$ is significantly greater than in the periodic driving scenario. As $N$ rises, the charging advantage becomes more noticeable, suggesting that the LR interaction in conjunction with linear driving is especially effective at introducing energy into the system. For $N=8$, the $W_{max}$ of linear driving is more than twice as high as periodic driving in both the cases (see Fig.~\ref{fig7a}, Fig.~\ref{fig7b}). This implies that linear driving transfers energy into the system more efficiently when there are LR interactions. This is because of the non-periodic linear driving resonates better with the broader coupling range. For nearest-neighbour interaction with smaller number of spins ($N=4$) the value of $W_{max}$ is comparable with respect to periodic and linear driving (see Fig.~\ref{fig7a}). In contrast, for long-range (LR) interactions, periodic driving (see Fig.~\ref{fig7b}) results in significantly lower work deposition. Consequently, opting for periodic driving in an LR interaction would lead to poor charging efficiency. Energy transfer becomes very efficient in bigger spin systems with LR contacts, as seen by the quick increase in $W_{max}$ with $N$ under linear driving in Fig.~\ref{fig7b}. Because all the spins have coupled more widely in LR, more of the system can be influenced by external forcing, which increases work deposition.On the other hand, for NN, the trend is milder, indicating that energy deposition scales with $N$ is less sharply. This should be expected since NN interaction constrains the system's collective response to the driving field by limiting the ability of every spin to interact with more than its nearest neighbors. This should suggest that in the presence of inherently long-range interactions, a linear drive might be exploited to best tap these interactions for maximum energy absorption. Conversely, periodic driving may be favored for simplicity and controlled pattern of energy deposition in systems where the interaction is NN.

\subsection{Optimal work deposition in a many-body quantum battery under maximum anisotropy}

\begin{figure*}[htp]
		\centering
		\begin{subfigure}[b]{0.45\textwidth}
			\centering
			\includegraphics[width=\textwidth]{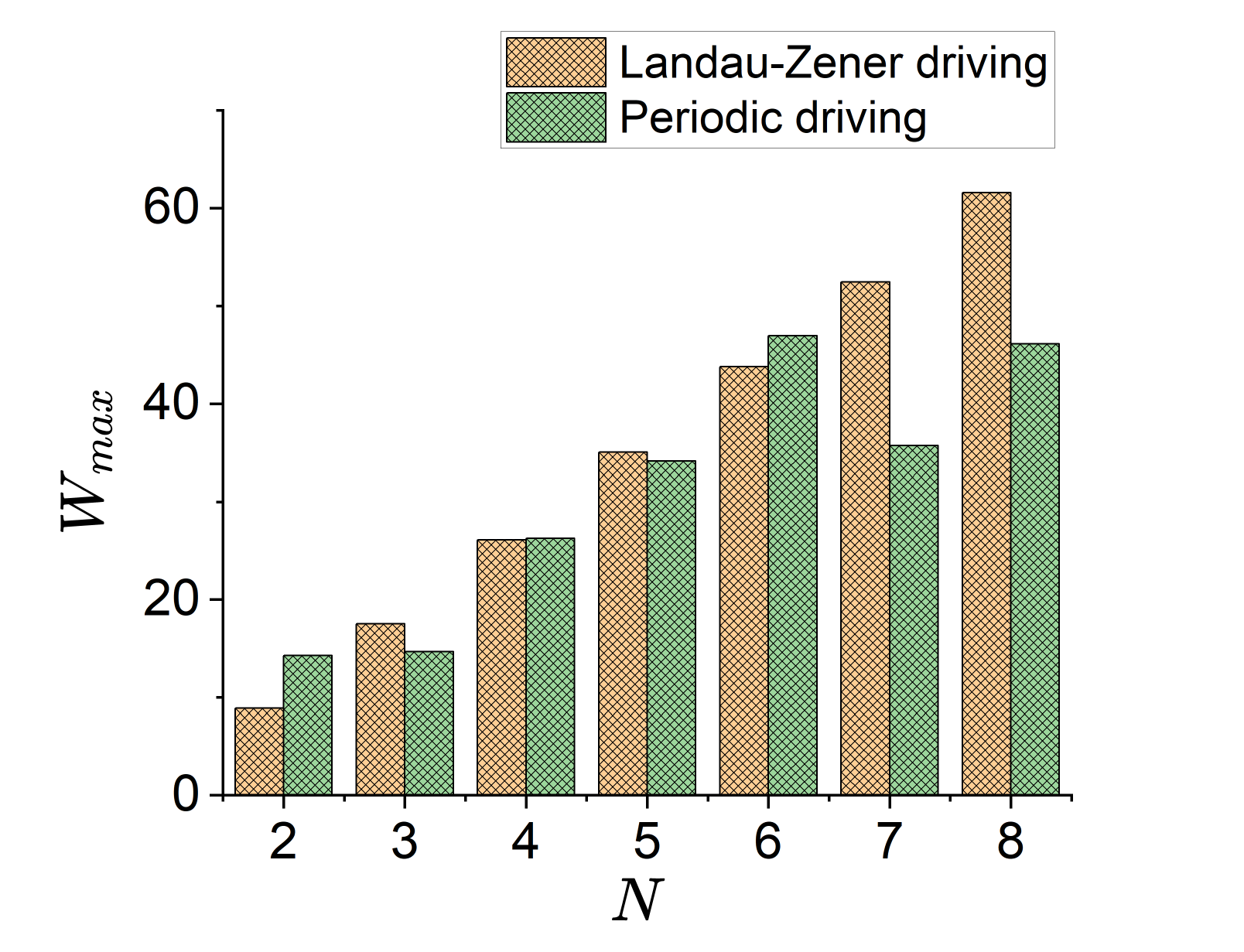}
			\caption{Nearest - Neighbour interaction}
			\label{fig8a}
		\end{subfigure}
		\hskip -0.0cm
		\begin{subfigure}[b]{0.45\textwidth}
			\centering
			\includegraphics[width=\textwidth]{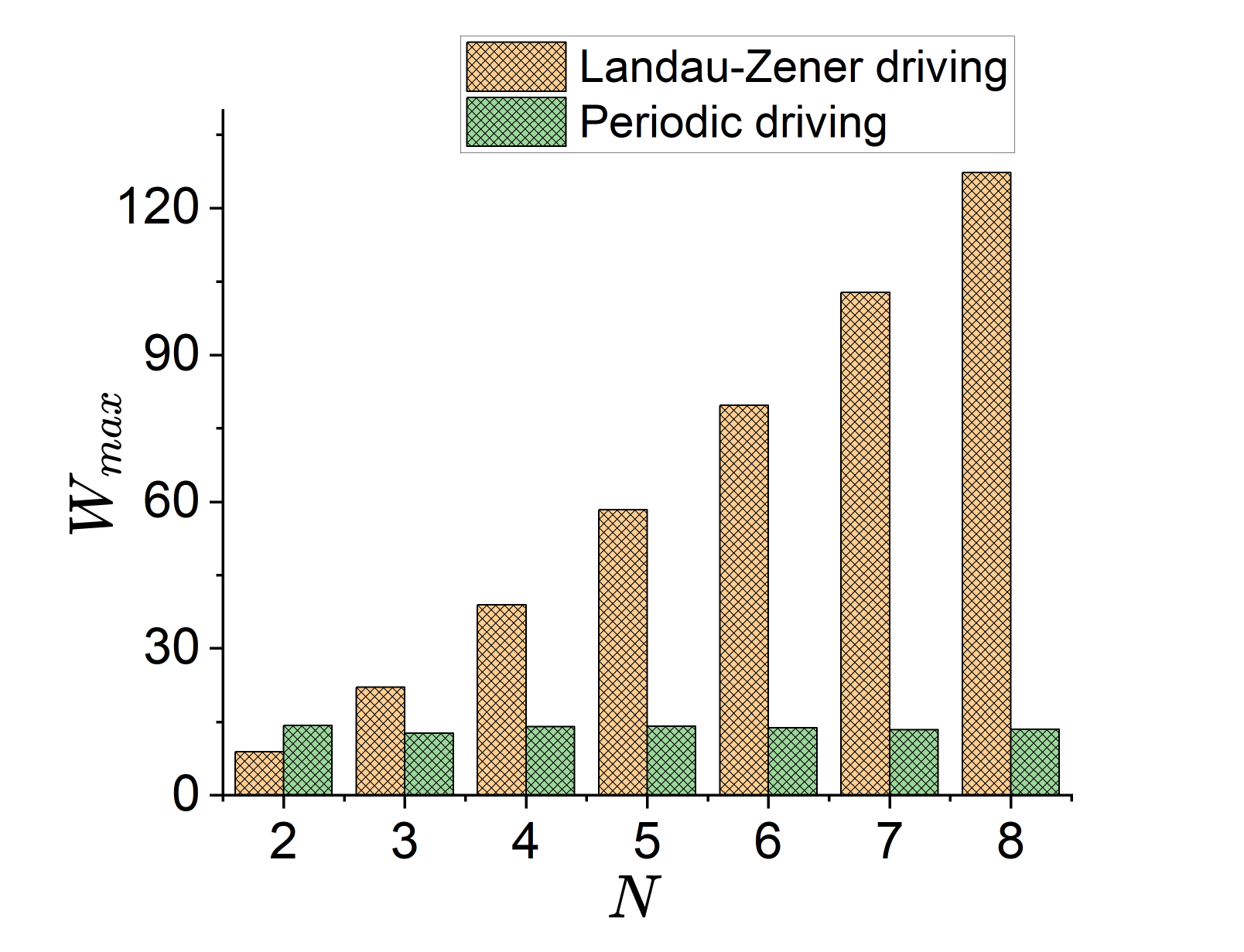}
			\caption{Long - range interaction}
			\label{fig8b}
		\end{subfigure}
		\caption{We have taken the maximum value of $\gamma$ and plot for the maximum value of $W$ in the over all time range $0 \leq \tau \leq 20$ for increasing values of the number of spins. $W_{max}$ Vs $N$ calculated for LR and NN. The parameters  are $g = 10B$, $\omega = 4B$, $\gamma = 1.0$ and $v = 10B$. The periodic driving and linear driving values are compared with respect to long-range (LR) and nearest-neighbour (NN) interaction.}
		\label{fig8}
	\end{figure*}

Next, we consider the maximum allowed value of the anisotropy to be $\gamma=1.0$, and conduct the comparative
analysis of the charging process in relation to both periodic and Landau-Zener driving. In Fig.~\ref{fig8}, we illustrate how the increase in the number of spins $N$ influences the maximum work deposited for both the nearest-neighbor interaction (see Fig.~\ref{fig8a}) and the long-range interaction (see Fig.~\ref{fig8b}) over a charging time interval of $0\leq B\tau \leq 20$. For increasing values of $N$, we represent the maximum work deposited $W_{max}$. Similar to Fig.~\ref{fig7}, the bar diagrams in Fig.~\ref{fig8a} and Fig.~\ref{fig8b} shows the nearest-neighbour and long-range interactions, respectively, for increasing values of $N$. The linear driving (LZ-driving) is shown by the brown column bars, while periodic driving is depicted by the green column bars. The many-body coupling strength is set to	$g=10B$, the anisotropy parameter to $\gamma=1.0$, and both the Landau-Zener driving and periodic driving parameters to $v = 10B$. The frequency of the periodic driving is fixed at $\omega = 4B$ for both the nearest-neighbor (Fig.~\ref{fig8a}) and long-range (Fig.~\ref{fig8b}) interaction scenarios. Up to $N=4$ in Fig.~\ref{fig8a}, both Landau-Zener and periodic driving produce comparable values of $W_{max}$, with only minor differences. This suggests that for small system sizes, the nature of the driving protocol (linear vs. periodic) has a limited impact on energy deposition when interactions are restricted to nearest neighbours. Starting from $N=5$, the Landau-Zener (linear) driving begins to show a more prominent increase in $W_{max}$ than the periodic driving. By $N=8$, the linear drive achieves a noticeably higher work deposition than the periodic drive, which indicates that the linear drive becomes more efficient in NN systems as the system size grows. The energy deposition increases rapidly with $N$ in Fig.~\ref{fig8b} for long-range interaction, and the bar heights indicate a robust enhancement in $W_{max}$ as $N$ increases. This implies that effective work deposition that increases significantly with system size is made possible by linear driving under LR interactions (see Fig.~\ref{fig8b}). Although periodic driving also causes $W_{max}$ to increase with $N$, the rate of increase is mild and plateaus at a particular value. For example, periodic driving only shows a slight increase in $W_{max}$ from $N=4$ to $N=8$, suggesting a saturation effect as $N$ increases. This implies that the long-range interaction does not play a significant role in cumulative energy deposition under periodic driving. It's evident that the continuous linear drive better utilizes the LR coupling than the periodic driving.
	
	From Fig.~\ref{fig8a}, on the contrary to the long-range example from Fig.~\ref{fig8b}, the maximum work deposition increases more gradually as the number of spins $N$ increases. The Landau-Zener and periodic driving types have comparable growth trends with $N$, but their magnitudes stay near for all values of $N$, with the differences being apparent at $N=6$ and above. As the number of spins $N$ increases, so does $W_{max}$ for both driving procedures. However, the linear driving case sees a significantly larger increase compared to periodic driving. We find that for $N=8$, $W_{max}$ for linear driving is more than three times that of periodic driving, indicating that as $N$ increases, the linear drive remains much more efficient in energy deposition in LR interactions (see Fig.~\ref{fig8b}). We found in NN systems from Fig.~\ref{fig8a}, the energy deposition trends suggest that for small systems (e.g., $N\leq4$), either driving protocol is suitable if comparable work deposition is the objective. The system behaves similarly under both protocols, making periodic driving a viable option for controlled, moderate energy input. For larger systems (e.g.,	$N>4$), linear driving becomes more advantageous for maximizing energy deposition, but the advantage is less drastic than in LR systems. Therefore, while linear driving is still slightly superior for higher energy input, periodic driving remains relatively efficient in NN systems (see Fig.~\ref{fig8a}). We found that given LR interaction conditions in Fig.~\ref{fig8b}, linear driving is the best option for applications where optimizing energy deposition is essential, including quantum information processing or designed energy transfer systems (like quantum batteries). The findings demonstrate that linear driving produces a large and scalable energy input, which is especially advantageous for bigger systems. Even while periodic driving is less efficient in terms of $W_{max}$, it may still be useful in circumstances where a moderate, controlled energy input is required because it seems to provide a more steady, less abrupt increase with system size. The Periodic driving provides steady and moderate energy dynamics that scale with system size for NN interaction applications where regulated energy deposition is sought. In situations like some quantum information processing methods that need consistent energy distribution without runaway accumulation, this might be advantageous \cite{polkovnikov2011colloquium,heyl2013dynamical}. The small advantage of linear driving in larger NN systems is less pronounced than in LR systems, but it might be useful in jobs that call for maximal energy transfer or excitation in extended spin chains. For systems where energy control is essential, like in quantum thermodynamics or thermal management in quantum devices, understanding the effects of interaction range and driving protocols can guide the selection of driving modes. Linear driving is beneficial when high energy absorption is the goal, while periodic driving could serve in applications needing steady-state or controlled energy dynamics. In larger systems, the linear driving’s high energy input could be harnessed for collective excitations or achieving critical behavior in LR-coupled spin systems, enhancing functionality in quantum simulations or other many-body applications.

\subsection{Maximum energy deposition in the quantum battery under strong driving}

\begin{figure*}[htp]
	\centering
	\begin{subfigure}[b]{0.45\textwidth}
		\centering
		\includegraphics[width=\textwidth]{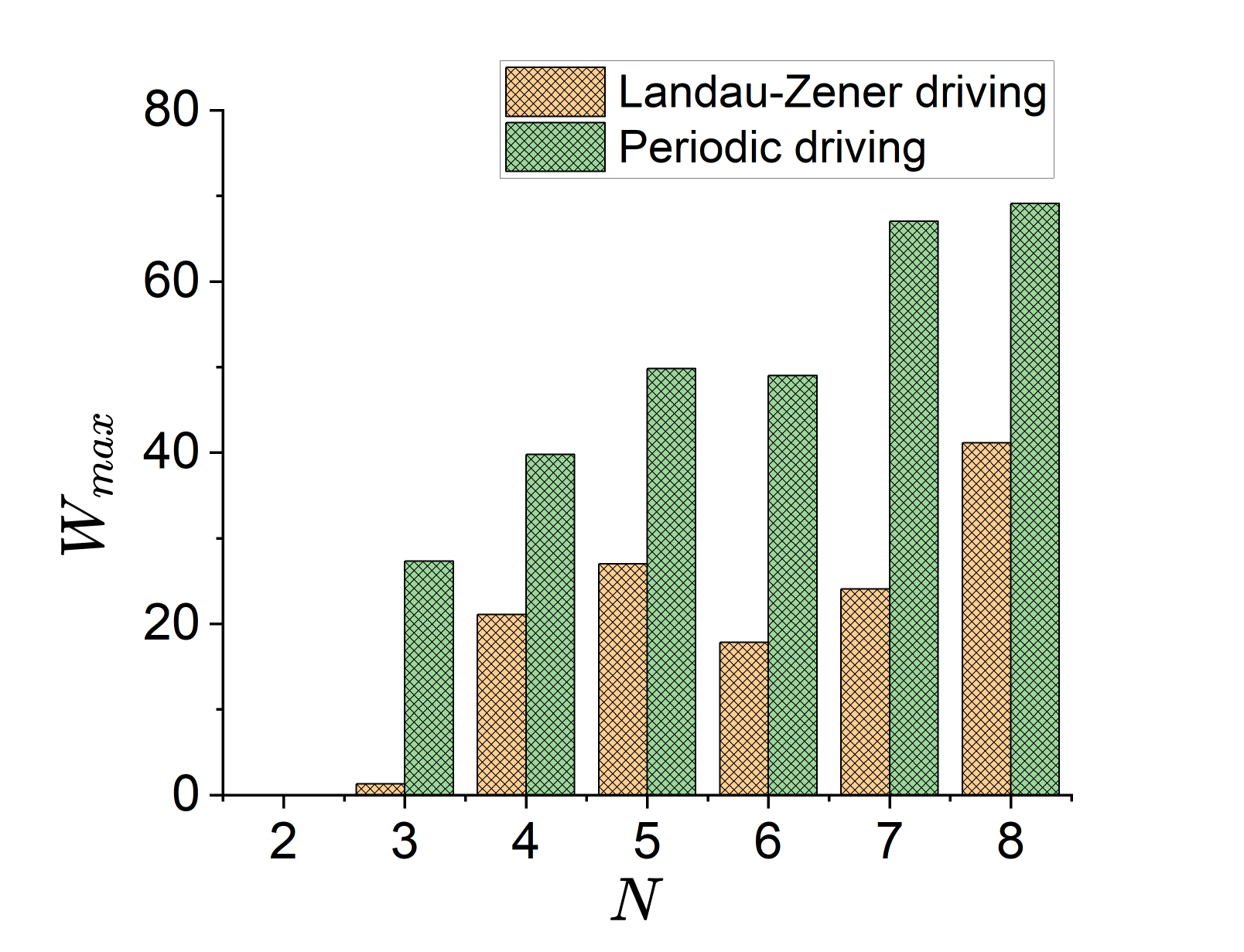}
		\caption{Nearest - Neighbour interaction}
		\label{fig9a}
	\end{subfigure}
	\hskip -0.0cm
	\begin{subfigure}[b]{0.45\textwidth}
		\centering
		\includegraphics[width=\textwidth]{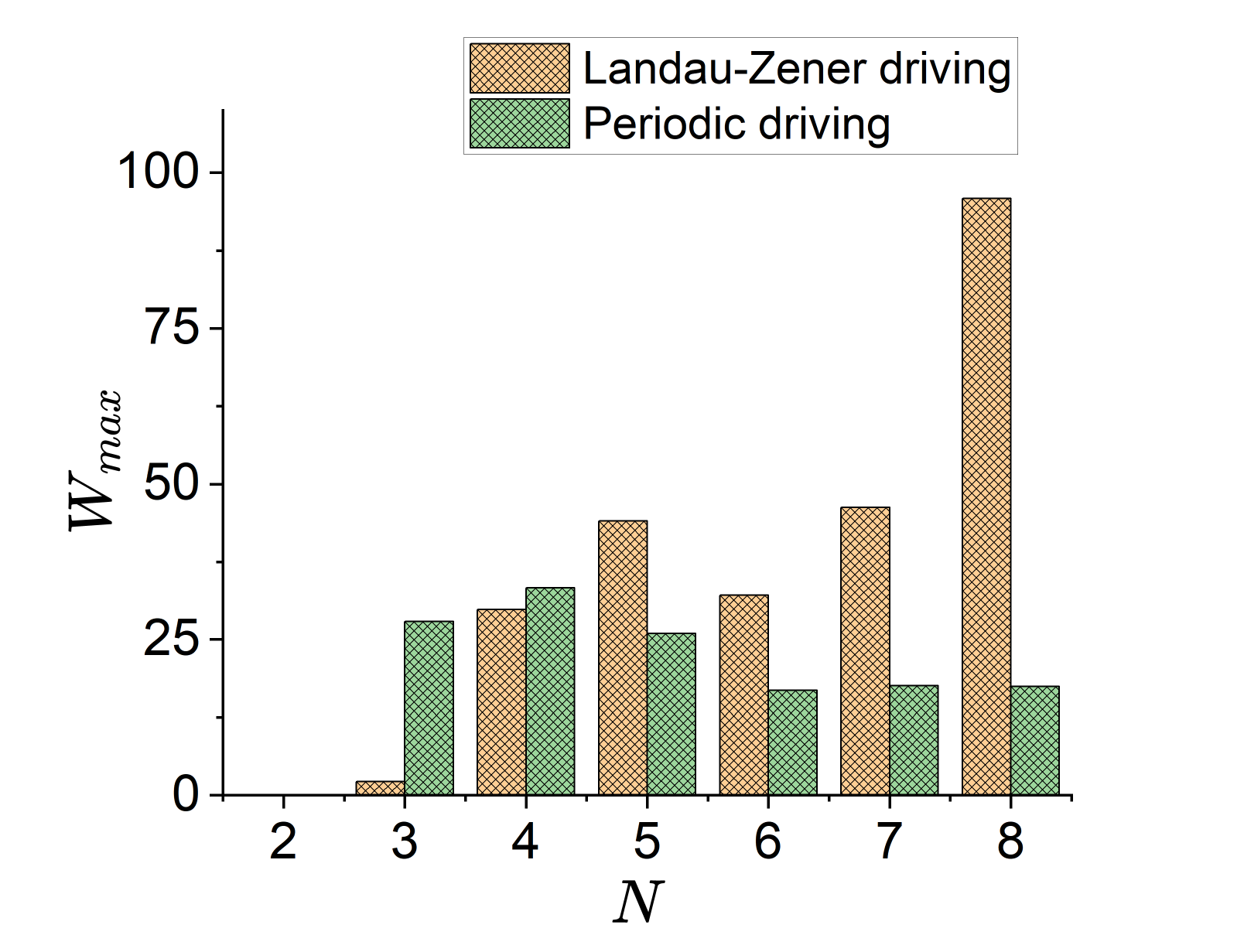}
		\caption{Long-range interaction}
		\label{fig9b}
	\end{subfigure}
	\caption{We have taken the maximum value of $v$ and plotted for the maximum value of $W$ in the over all time range $0 \leq \tau \leq 20$ for increasing values of the number of spins. $W_{max}$ Vs $N$ calculated for LR and NN. The parameters  are $g = 10B$, $\omega = 4B$, $\gamma = 0.5$ and $v = 10B$. The periodic driving and linear driving values are compared with respect to long-range (LR) and nearest-neighbour (NN) interaction.}
	\label{fig9}
	\end{figure*}

In this subsection, we try to analyze the case where we take the maximum value (which we have assigned throughout this paper)
of the LZ-driving amplitude $v=10B$ and keep the rest of the parameters same, and compare the maximum work deposited in both
LZ-driving and periodic driving under the LR and NN interaction. Both plots (see Fig.~\ref{fig9}) represent the maximum work deposited ($W_{max}$) in a spin-chain quantum battery with varying system size $N$ under two types of driving mechanisms: Landau-Zener driving and Periodic driving. The maximum work deposited for the NN interaction (Fig.~\ref{fig9a}) and the LR interaction (Fig.~\ref{fig9b}) during a charging time interval of $0\leq B\tau \leq 20$ is affected by the increase in the number of spins $N$. We indicate the maximum work deposited as $W_{max}$ for increasing values of $N$. The bar graphs in Fig.~\ref{fig9a} and Fig.~\ref{fig9b}, depict the nearest-neighbor and long-range interactions respectively, for increasing values of $N$ where brown column bar graph represents linear driving and the green column bar graph represents periodic driving. The parameters are set to $\gamma = 0.5$ (anisotropy parameter), and the many-body coupling strength to $g=10B$. For both the long-range (Fig.~\ref{fig9b}) and nearest-neighbor (Fig.~\ref{fig9a}) interaction situations, the frequency of the periodic driving is set at $\omega = 4B$.

	In the NN interaction from Fig.~\ref{fig9a}, both driving types show a fairly steady increase in $W_{max}$ with $N$. The trend reverses between the two driving mechanisms. Here, Periodic driving deposits more work across all $N$, with a steady growth. For $N=8$, $W_{max}$ peaks at around $70$. Landau-Zener driving shows a modest increase with $N$, with values consistently below $W_{max} \sim 50$. This indicating a more restrained energy transfer due to the restricted range of spin interactions. The work deposition increases with $N$ but at a slower rate in Fig.~\ref{fig9a} compared to the LR configuration in Fig.~\ref{fig9b}. This milder growth pattern suggests that the energy transfer in LZ is not as efficient when limited to NN interactions, as the system has fewer pathways for distributing energy across spins. Periodic driving exhibits a more efficient work deposition here, consistently surpassing LZ driving in $W_{max}$ as $N$ grows. This might indicate that periodic oscillations resonate better with the local, nearest-neighbor couplings, allowing energy to be distributed more efficiently in this configuration. In the nearest-neighbour case (refer Fig.~\ref{fig9a}), periodic driving seems to leverage the NN configuration effectively, with a more regular increase in $W_{max}$. It also becomes the more dominant mechanism for energy deposition as $N$ increases. Nearest-neighbor interactions favor periodic driving, potentially due to resonance effects and localized energy transfer mechanisms that match well with periodic modulation. We observe from Fig.~\ref{fig9b} that for the long-range case, $W_{max}$ generally increases with $N$, but with noticeable differences in the rate and pattern of increase between the two driving protocols. On further observation from Fig.~\ref{fig9b} the work deposition for LZ driving increases more consistently with $N$ than in previous cases. For Landau-Zener driving, $W_{max}$	exhibits a dramatic increase with $N$, particularly for $N\geq7$. The growth is non-linear and sharply peaks at $N=8$, reaching over $W_{max} \sim 100$. For Periodic driving, $W_{max}$ increases gradually with $N$, with a less pronounced jump at $N=7$ and $N=8$, staying below $W_{max}\sim30$. Landau-Zener driving significantly outperforms periodic driving, especially for larger $N$, suggesting a stronger enhancement of work deposition via non-local couplings. The increase becomes more pronounced, especially as $N$ approaches 8, with a large spike in $W_{max}$. This jump suggests that larger system sizes benefit from the adiabatic-like transition characteristic of LZ driving, as the system can absorb energy more effectively over time. In periodic driving, $W_{max}$ shows a similar overall increasing trend with $N$, but the rate of increase is slightly more gradual compared to LZ driving. This indicates that, while periodic driving deposits significant energy, it doesn’t amplify energy transfer as strongly with system size in this configuration. Our further investigation suggests that in Fig.~\ref{fig9b} for smaller system sizes (especially $N<5$), periodic driving seems more efficient, depositing slightly more work compared to LZ. However, as $N$ increases, LZ driving becomes more efficient, suggesting that its adiabatic energy transfer mechanism works better for larger systems in the LR case. The non-linear growth of $W_{max}$ under Landau-Zener driving (with long-range interactions) suggests a quantum many-body effect, where the cooperative dynamics of the spins enhance energy deposition. For bigger systems, LZ driving works better, because of the wider range of interactions that enhance work deposition. Landau-Zener driving demonstrates increased sensitivity to the interaction range, exhibiting a significant advantage in long-range configurations but displaying limited efficacy in nearest-neighbor setups. Conversely, periodic driving shows robustness and consistency, delivering superior performance in nearest-neighbor configurations while maintaining moderate work values in long-range interactions. The higher spin count in LR setups is more advantageous for this drive. For NN interactions, periodic driving works best, depositing more energy as N rises. This demonstrates that localized interaction schemes, where energy resonance between closely interacting spins is more effective, are highly compatible with periodic driving. Periodic driving’s smoother scaling with $N$ across both interaction types indicates a more uniform energy deposition process, irrespective of the interaction range. Long-range interactions significantly amplify the deposited work under Landau-Zener driving, due to enhanced correlations and energy transfer over extended distances in the spin chain. Nearest-neighbor interactions favor periodic driving, potentially due to resonance effects and localized energy transfer mechanisms that match well with periodic modulation.

	\section{conclusions}\label{sec:conclusions}

In this work, we have investigated the charging dynamics of a many-body quantum battery, powered by a Landau-Zener field and modeled as a Heisenberg XY spin chain with $N$ interacting spin-1/2 particles under an external magnetic field. By comparing the Landau-Zener and periodic driving mechanisms, we observe distinct variations in energy deposition and storage efficiency across both the long-range and nearest-neighbor interaction regimes. It turns out that the Landau-Zener driving mode can significantly enhance maximum work deposition, particularly as the system size increases, highlighting the effectiveness of non-periodic driving fields in optimizing energy storage. In the long-range interaction regime, the maximum work deposition scales almost linearly with increasing system size, suggesting that long-range interactions can enhance energy storage efficiency as the chain elongates. This way, quantum spin chains with long-range interactions present promising options for scalable quantum battery systems. However, the saturation trend in maximum work deposition for the nearest-neighbor configuration highlights the limitations of short-range interactions in storing energy within extended spin systems. Our work further suggests that the performance of a quantum battery can be significantly improved by appropriately tuning the driving field and interaction strength. For smaller system sizes, periodic driving surpasses Landau-Zener driving, highlighting its potential for energy storage in localized, small-scale quantum systems where maintaining long-range coherence is challenging. Nonetheless, our work demonstrates that a one-dimensional spin chain can serve as an effective model for a quantum battery, even with a linearly time-varying charging field. This framework could be extended in future works to explore the effects of decoherence in these systems while investigating alternative spin configurations, coupling strategies, and dynamic driving fields to further enhance quantum battery performance.
	
	\section*{Acknowledgments}
	AS and SAS would like to thank the Science and Engineering Research Board (SERB), Government of India via Project No. CRG/2022/007836 for providing necessary computational resources through the SERB-funded GPGPU system at SRMIST, India. We further thank the High-Performance Computing Center at SRMIST to facilitate our work.

	\bibliographystyle{apsrev}
	\bibliography{ReferenceBattery}

\begin{thebibliography}{49}
\expandafter\ifx\csname natexlab\endcsname\relax\def\natexlab#1{#1}\fi
\expandafter\ifx\csname bibnamefont\endcsname\relax
  \def\bibnamefont#1{#1}\fi
\expandafter\ifx\csname bibfnamefont\endcsname\relax
  \def\bibfnamefont#1{#1}\fi
\expandafter\ifx\csname citenamefont\endcsname\relax
  \def\citenamefont#1{#1}\fi
\expandafter\ifx\csname url\endcsname\relax
  \def\url#1{\texttt{#1}}\fi
\expandafter\ifx\csname urlprefix\endcsname\relax\def\urlprefix{URL }\fi
\providecommand{\bibinfo}[2]{#2}
\providecommand{\eprint}[2][]{\url{#2}}

\bibitem[{\citenamefont{Nielsen and Chuang}(2010)}]{nielsen2010quantum}
\bibinfo{author}{\bibfnamefont{M.~A.} \bibnamefont{Nielsen}} \bibnamefont{and}
  \bibinfo{author}{\bibfnamefont{I.~L.} \bibnamefont{Chuang}},
  \emph{\bibinfo{title}{Quantum computation and quantum information}}
  (\bibinfo{publisher}{Cambridge University Press}, \bibinfo{year}{2010}).

\bibitem[{\citenamefont{Xu et~al.}(2020)\citenamefont{Xu, Ma, Zhang, Lo, and
  Pan}}]{xu2020secure}
\bibinfo{author}{\bibfnamefont{F.}~\bibnamefont{Xu}},
  \bibinfo{author}{\bibfnamefont{X.}~\bibnamefont{Ma}},
  \bibinfo{author}{\bibfnamefont{Q.}~\bibnamefont{Zhang}},
  \bibinfo{author}{\bibfnamefont{H.-K.} \bibnamefont{Lo}}, \bibnamefont{and}
  \bibinfo{author}{\bibfnamefont{J.-W.} \bibnamefont{Pan}},
  \bibinfo{journal}{Reviews of modern physics} \textbf{\bibinfo{volume}{92}},
  \bibinfo{pages}{025002} (\bibinfo{year}{2020}).

\bibitem[{\citenamefont{Campaioli et~al.}(2018)\citenamefont{Campaioli,
  Pollock, and Vinjanampathy}}]{campaioli2018quantum}
\bibinfo{author}{\bibfnamefont{F.}~\bibnamefont{Campaioli}},
  \bibinfo{author}{\bibfnamefont{F.~A.} \bibnamefont{Pollock}},
  \bibnamefont{and}
  \bibinfo{author}{\bibfnamefont{S.}~\bibnamefont{Vinjanampathy}},
  \bibinfo{journal}{Thermodynamics in the Quantum Regime: Fundamental Aspects
  and New Directions} pp. \bibinfo{pages}{207--225} (\bibinfo{year}{2018}).

\bibitem[{\citenamefont{Quach et~al.}(2023)\citenamefont{Quach, Cerullo, and
  Virgili}}]{quach2023quantum}
\bibinfo{author}{\bibfnamefont{J.~Q.} \bibnamefont{Quach}},
  \bibinfo{author}{\bibfnamefont{G.}~\bibnamefont{Cerullo}}, \bibnamefont{and}
  \bibinfo{author}{\bibfnamefont{T.}~\bibnamefont{Virgili}},
  \bibinfo{journal}{Joule} \textbf{\bibinfo{volume}{7}}, \bibinfo{pages}{2195}
  (\bibinfo{year}{2023}).

\bibitem[{\citenamefont{Campaioli et~al.}(2024)\citenamefont{Campaioli,
  Gherardini, Quach, Polini, and Andolina}}]{campaioli2024colloquium}
\bibinfo{author}{\bibfnamefont{F.}~\bibnamefont{Campaioli}},
  \bibinfo{author}{\bibfnamefont{S.}~\bibnamefont{Gherardini}},
  \bibinfo{author}{\bibfnamefont{J.~Q.} \bibnamefont{Quach}},
  \bibinfo{author}{\bibfnamefont{M.}~\bibnamefont{Polini}}, \bibnamefont{and}
  \bibinfo{author}{\bibfnamefont{G.~M.} \bibnamefont{Andolina}},
  \bibinfo{journal}{Reviews of Modern Physics} \textbf{\bibinfo{volume}{96}},
  \bibinfo{pages}{031001} (\bibinfo{year}{2024}).

\bibitem[{\citenamefont{Xiang and Guo}(2013)}]{xiang2013quantum}
\bibinfo{author}{\bibfnamefont{G.-Y.} \bibnamefont{Xiang}} \bibnamefont{and}
  \bibinfo{author}{\bibfnamefont{G.-C.} \bibnamefont{Guo}},
  \bibinfo{journal}{Chinese Physics B} \textbf{\bibinfo{volume}{22}},
  \bibinfo{pages}{110601} (\bibinfo{year}{2013}).

\bibitem[{\citenamefont{Alicki and Fannes}(2013)}]{alicki2013entanglement}
\bibinfo{author}{\bibfnamefont{R.}~\bibnamefont{Alicki}} \bibnamefont{and}
  \bibinfo{author}{\bibfnamefont{M.}~\bibnamefont{Fannes}},
  \bibinfo{journal}{Physical Review E} \textbf{\bibinfo{volume}{87}},
  \bibinfo{pages}{042123} (\bibinfo{year}{2013}).

\bibitem[{\citenamefont{Ferraro et~al.}(2018)\citenamefont{Ferraro, Campisi,
  Andolina, Pellegrini, and Polini}}]{ferraro2018high}
\bibinfo{author}{\bibfnamefont{D.}~\bibnamefont{Ferraro}},
  \bibinfo{author}{\bibfnamefont{M.}~\bibnamefont{Campisi}},
  \bibinfo{author}{\bibfnamefont{G.~M.} \bibnamefont{Andolina}},
  \bibinfo{author}{\bibfnamefont{V.}~\bibnamefont{Pellegrini}},
  \bibnamefont{and} \bibinfo{author}{\bibfnamefont{M.}~\bibnamefont{Polini}},
  \bibinfo{journal}{Physical review letters} \textbf{\bibinfo{volume}{120}},
  \bibinfo{pages}{117702} (\bibinfo{year}{2018}).

\bibitem[{\citenamefont{Rossini et~al.}(2020)\citenamefont{Rossini, Andolina,
  Rosa, Carrega, and Polini}}]{rossini2020quantum}
\bibinfo{author}{\bibfnamefont{D.}~\bibnamefont{Rossini}},
  \bibinfo{author}{\bibfnamefont{G.~M.} \bibnamefont{Andolina}},
  \bibinfo{author}{\bibfnamefont{D.}~\bibnamefont{Rosa}},
  \bibinfo{author}{\bibfnamefont{M.}~\bibnamefont{Carrega}}, \bibnamefont{and}
  \bibinfo{author}{\bibfnamefont{M.}~\bibnamefont{Polini}},
  \bibinfo{journal}{Physical Review Letters} \textbf{\bibinfo{volume}{125}},
  \bibinfo{pages}{236402} (\bibinfo{year}{2020}).

\bibitem[{\citenamefont{Xu et~al.}(2021)\citenamefont{Xu, Zhu, Zhang, and
  Liu}}]{xu2021enhancing}
\bibinfo{author}{\bibfnamefont{K.}~\bibnamefont{Xu}},
  \bibinfo{author}{\bibfnamefont{H.-J.} \bibnamefont{Zhu}},
  \bibinfo{author}{\bibfnamefont{G.-F.} \bibnamefont{Zhang}}, \bibnamefont{and}
  \bibinfo{author}{\bibfnamefont{W.-M.} \bibnamefont{Liu}},
  \bibinfo{journal}{Physical Review E} \textbf{\bibinfo{volume}{104}},
  \bibinfo{pages}{064143} (\bibinfo{year}{2021}).

\bibitem[{\citenamefont{Seah et~al.}(2021)\citenamefont{Seah, Perarnau-Llobet,
  Haack, Brunner, and Nimmrichter}}]{seah2021quantum}
\bibinfo{author}{\bibfnamefont{S.}~\bibnamefont{Seah}},
  \bibinfo{author}{\bibfnamefont{M.}~\bibnamefont{Perarnau-Llobet}},
  \bibinfo{author}{\bibfnamefont{G.}~\bibnamefont{Haack}},
  \bibinfo{author}{\bibfnamefont{N.}~\bibnamefont{Brunner}}, \bibnamefont{and}
  \bibinfo{author}{\bibfnamefont{S.}~\bibnamefont{Nimmrichter}},
  \bibinfo{journal}{Physical Review Letters} \textbf{\bibinfo{volume}{127}},
  \bibinfo{pages}{100601} (\bibinfo{year}{2021}).

\bibitem[{\citenamefont{Campaioli et~al.}(2017)\citenamefont{Campaioli,
  Pollock, Binder, C{\'e}leri, Goold, Vinjanampathy, and
  Modi}}]{campaioli2017enhancing}
\bibinfo{author}{\bibfnamefont{F.}~\bibnamefont{Campaioli}},
  \bibinfo{author}{\bibfnamefont{F.~A.} \bibnamefont{Pollock}},
  \bibinfo{author}{\bibfnamefont{F.~C.} \bibnamefont{Binder}},
  \bibinfo{author}{\bibfnamefont{L.}~\bibnamefont{C{\'e}leri}},
  \bibinfo{author}{\bibfnamefont{J.}~\bibnamefont{Goold}},
  \bibinfo{author}{\bibfnamefont{S.}~\bibnamefont{Vinjanampathy}},
  \bibnamefont{and} \bibinfo{author}{\bibfnamefont{K.}~\bibnamefont{Modi}},
  \bibinfo{journal}{Physical review letters} \textbf{\bibinfo{volume}{118}},
  \bibinfo{pages}{150601} (\bibinfo{year}{2017}).

\bibitem[{\citenamefont{Ghosh et~al.}(2020)\citenamefont{Ghosh, Chanda, and
  Sen}}]{ghosh2020enhancement}
\bibinfo{author}{\bibfnamefont{S.}~\bibnamefont{Ghosh}},
  \bibinfo{author}{\bibfnamefont{T.}~\bibnamefont{Chanda}}, \bibnamefont{and}
  \bibinfo{author}{\bibfnamefont{A.}~\bibnamefont{Sen}},
  \bibinfo{journal}{Physical Review A} \textbf{\bibinfo{volume}{101}},
  \bibinfo{pages}{032115} (\bibinfo{year}{2020}).

\bibitem[{\citenamefont{Andolina et~al.}(2019)\citenamefont{Andolina, Keck,
  Mari, Campisi, Giovannetti, and Polini}}]{andolina2019extractable}
\bibinfo{author}{\bibfnamefont{G.~M.} \bibnamefont{Andolina}},
  \bibinfo{author}{\bibfnamefont{M.}~\bibnamefont{Keck}},
  \bibinfo{author}{\bibfnamefont{A.}~\bibnamefont{Mari}},
  \bibinfo{author}{\bibfnamefont{M.}~\bibnamefont{Campisi}},
  \bibinfo{author}{\bibfnamefont{V.}~\bibnamefont{Giovannetti}},
  \bibnamefont{and} \bibinfo{author}{\bibfnamefont{M.}~\bibnamefont{Polini}},
  \bibinfo{journal}{Physical review letters} \textbf{\bibinfo{volume}{122}},
  \bibinfo{pages}{047702} (\bibinfo{year}{2019}).

\bibitem[{\citenamefont{Bhattacharjee and
  Dutta}(2021)}]{bhattacharjee2021quantum}
\bibinfo{author}{\bibfnamefont{S.}~\bibnamefont{Bhattacharjee}}
  \bibnamefont{and} \bibinfo{author}{\bibfnamefont{A.}~\bibnamefont{Dutta}},
  \bibinfo{journal}{The European Physical Journal B}
  \textbf{\bibinfo{volume}{94}}, \bibinfo{pages}{1} (\bibinfo{year}{2021}).

\bibitem[{\citenamefont{Cangemi et~al.}(2024)\citenamefont{Cangemi, Bhadra, and
  Levy}}]{cangemi2024quantum}
\bibinfo{author}{\bibfnamefont{L.~M.} \bibnamefont{Cangemi}},
  \bibinfo{author}{\bibfnamefont{C.}~\bibnamefont{Bhadra}}, \bibnamefont{and}
  \bibinfo{author}{\bibfnamefont{A.}~\bibnamefont{Levy}},
  \bibinfo{journal}{Physics Reports} \textbf{\bibinfo{volume}{1087}},
  \bibinfo{pages}{1} (\bibinfo{year}{2024}).

\bibitem[{\citenamefont{Le et~al.}(2018)\citenamefont{Le, Levinsen, Modi,
  Parish, and Pollock}}]{le2018spin}
\bibinfo{author}{\bibfnamefont{T.~P.} \bibnamefont{Le}},
  \bibinfo{author}{\bibfnamefont{J.}~\bibnamefont{Levinsen}},
  \bibinfo{author}{\bibfnamefont{K.}~\bibnamefont{Modi}},
  \bibinfo{author}{\bibfnamefont{M.~M.} \bibnamefont{Parish}},
  \bibnamefont{and} \bibinfo{author}{\bibfnamefont{F.~A.}
  \bibnamefont{Pollock}}, \bibinfo{journal}{Physical Review A}
  \textbf{\bibinfo{volume}{97}}, \bibinfo{pages}{022106}
  (\bibinfo{year}{2018}).

\bibitem[{\citenamefont{Hadipour et~al.}(2023)\citenamefont{Hadipour, Haseli,
  Dolatkhah, and Rashidi}}]{hadipour2023study}
\bibinfo{author}{\bibfnamefont{M.}~\bibnamefont{Hadipour}},
  \bibinfo{author}{\bibfnamefont{S.}~\bibnamefont{Haseli}},
  \bibinfo{author}{\bibfnamefont{H.}~\bibnamefont{Dolatkhah}},
  \bibnamefont{and} \bibinfo{author}{\bibfnamefont{M.}~\bibnamefont{Rashidi}},
  \bibinfo{journal}{Scientific Reports} \textbf{\bibinfo{volume}{13}},
  \bibinfo{pages}{10672} (\bibinfo{year}{2023}).

\bibitem[{\citenamefont{{\v{S}}afr{\'a}nek
  et~al.}(2023)\citenamefont{{\v{S}}afr{\'a}nek, Rosa, and
  Binder}}]{vsafranek2023work}
\bibinfo{author}{\bibfnamefont{D.}~\bibnamefont{{\v{S}}afr{\'a}nek}},
  \bibinfo{author}{\bibfnamefont{D.}~\bibnamefont{Rosa}}, \bibnamefont{and}
  \bibinfo{author}{\bibfnamefont{F.~C.} \bibnamefont{Binder}},
  \bibinfo{journal}{Physical Review Letters} \textbf{\bibinfo{volume}{130}},
  \bibinfo{pages}{210401} (\bibinfo{year}{2023}).

\bibitem[{\citenamefont{Mitchison et~al.}(2021)\citenamefont{Mitchison, Goold,
  and Prior}}]{mitchison2021charging}
\bibinfo{author}{\bibfnamefont{M.~T.} \bibnamefont{Mitchison}},
  \bibinfo{author}{\bibfnamefont{J.}~\bibnamefont{Goold}}, \bibnamefont{and}
  \bibinfo{author}{\bibfnamefont{J.}~\bibnamefont{Prior}},
  \bibinfo{journal}{Quantum} \textbf{\bibinfo{volume}{5}}, \bibinfo{pages}{500}
  (\bibinfo{year}{2021}).

\bibitem[{\citenamefont{Downing and Ukhtary}(2023)}]{downing2023quantum}
\bibinfo{author}{\bibfnamefont{C.~A.} \bibnamefont{Downing}} \bibnamefont{and}
  \bibinfo{author}{\bibfnamefont{M.~S.} \bibnamefont{Ukhtary}},
  \bibinfo{journal}{Communications Physics} \textbf{\bibinfo{volume}{6}},
  \bibinfo{pages}{322} (\bibinfo{year}{2023}).

\bibitem[{\citenamefont{Liu et~al.}(2019)\citenamefont{Liu, Segal, and
  Hanna}}]{liu2019loss}
\bibinfo{author}{\bibfnamefont{J.}~\bibnamefont{Liu}},
  \bibinfo{author}{\bibfnamefont{D.}~\bibnamefont{Segal}}, \bibnamefont{and}
  \bibinfo{author}{\bibfnamefont{G.}~\bibnamefont{Hanna}},
  \bibinfo{journal}{The Journal of Physical Chemistry C}
  \textbf{\bibinfo{volume}{123}}, \bibinfo{pages}{18303}
  (\bibinfo{year}{2019}).

\bibitem[{\citenamefont{Catalano et~al.}(2023)\citenamefont{Catalano,
  Giampaolo, Morsch, Giovannetti, and Franchini}}]{catalano2023frustrating}
\bibinfo{author}{\bibfnamefont{A.~G.} \bibnamefont{Catalano}},
  \bibinfo{author}{\bibfnamefont{S.~M.} \bibnamefont{Giampaolo}},
  \bibinfo{author}{\bibfnamefont{O.}~\bibnamefont{Morsch}},
  \bibinfo{author}{\bibfnamefont{V.}~\bibnamefont{Giovannetti}},
  \bibnamefont{and}
  \bibinfo{author}{\bibfnamefont{F.}~\bibnamefont{Franchini}},
  \bibinfo{journal}{arXiv preprint arXiv:2307.02529}  (\bibinfo{year}{2023}).

\bibitem[{\citenamefont{Jin and Ye}(2011)}]{jin2011polar}
\bibinfo{author}{\bibfnamefont{D.~S.} \bibnamefont{Jin}} \bibnamefont{and}
  \bibinfo{author}{\bibfnamefont{J.}~\bibnamefont{Ye}},
  \bibinfo{journal}{Physics Today} \textbf{\bibinfo{volume}{64}},
  \bibinfo{pages}{27} (\bibinfo{year}{2011}).

\bibitem[{\citenamefont{Lewenstein et~al.}(2007)\citenamefont{Lewenstein,
  Sanpera, Ahufinger, Damski, Sen, and Sen}}]{lewenstein2007ultracold}
\bibinfo{author}{\bibfnamefont{M.}~\bibnamefont{Lewenstein}},
  \bibinfo{author}{\bibfnamefont{A.}~\bibnamefont{Sanpera}},
  \bibinfo{author}{\bibfnamefont{V.}~\bibnamefont{Ahufinger}},
  \bibinfo{author}{\bibfnamefont{B.}~\bibnamefont{Damski}},
  \bibinfo{author}{\bibfnamefont{A.}~\bibnamefont{Sen}}, \bibnamefont{and}
  \bibinfo{author}{\bibfnamefont{U.}~\bibnamefont{Sen}},
  \bibinfo{journal}{Advances in Physics} \textbf{\bibinfo{volume}{56}},
  \bibinfo{pages}{243} (\bibinfo{year}{2007}).

\bibitem[{\citenamefont{Andolina et~al.}(2018)\citenamefont{Andolina, Farina,
  Mari, Pellegrini, Giovannetti, and Polini}}]{andolina2018charger}
\bibinfo{author}{\bibfnamefont{G.~M.} \bibnamefont{Andolina}},
  \bibinfo{author}{\bibfnamefont{D.}~\bibnamefont{Farina}},
  \bibinfo{author}{\bibfnamefont{A.}~\bibnamefont{Mari}},
  \bibinfo{author}{\bibfnamefont{V.}~\bibnamefont{Pellegrini}},
  \bibinfo{author}{\bibfnamefont{V.}~\bibnamefont{Giovannetti}},
  \bibnamefont{and} \bibinfo{author}{\bibfnamefont{M.}~\bibnamefont{Polini}},
  \bibinfo{journal}{Physical Review B} \textbf{\bibinfo{volume}{98}},
  \bibinfo{pages}{205423} (\bibinfo{year}{2018}).

\bibitem[{\citenamefont{Grazi et~al.}(2024)\citenamefont{Grazi, Shaikh,
  Sassetti, Ziani, and Ferraro}}]{grazi2024enhancing}
\bibinfo{author}{\bibfnamefont{R.}~\bibnamefont{Grazi}},
  \bibinfo{author}{\bibfnamefont{D.~S.} \bibnamefont{Shaikh}},
  \bibinfo{author}{\bibfnamefont{M.}~\bibnamefont{Sassetti}},
  \bibinfo{author}{\bibfnamefont{N.~T.} \bibnamefont{Ziani}}, \bibnamefont{and}
  \bibinfo{author}{\bibfnamefont{D.}~\bibnamefont{Ferraro}},
  \bibinfo{journal}{arXiv preprint arXiv:2402.09169}  (\bibinfo{year}{2024}).

\bibitem[{\citenamefont{Du et~al.}(2024)\citenamefont{Du, Chen, Wang, Yu, Guo,
  Qu, and Zhang}}]{du2024quantum}
\bibinfo{author}{\bibfnamefont{Y.}~\bibnamefont{Du}},
  \bibinfo{author}{\bibfnamefont{W.}~\bibnamefont{Chen}},
  \bibinfo{author}{\bibfnamefont{Y.}~\bibnamefont{Wang}},
  \bibinfo{author}{\bibfnamefont{Y.}~\bibnamefont{Yu}},
  \bibinfo{author}{\bibfnamefont{K.}~\bibnamefont{Guo}},
  \bibinfo{author}{\bibfnamefont{G.}~\bibnamefont{Qu}}, \bibnamefont{and}
  \bibinfo{author}{\bibfnamefont{J.}~\bibnamefont{Zhang}},
  \bibinfo{journal}{Nano-Micro Letters} \textbf{\bibinfo{volume}{16}},
  \bibinfo{pages}{100} (\bibinfo{year}{2024}).

\bibitem[{\citenamefont{Zueco et~al.}(2009)\citenamefont{Zueco, Galve, Kohler,
  and H{\"a}nggi}}]{zueco2009quantum}
\bibinfo{author}{\bibfnamefont{D.}~\bibnamefont{Zueco}},
  \bibinfo{author}{\bibfnamefont{F.}~\bibnamefont{Galve}},
  \bibinfo{author}{\bibfnamefont{S.}~\bibnamefont{Kohler}}, \bibnamefont{and}
  \bibinfo{author}{\bibfnamefont{P.}~\bibnamefont{H{\"a}nggi}},
  \bibinfo{journal}{Physical Review A} \textbf{\bibinfo{volume}{80}},
  \bibinfo{pages}{042303} (\bibinfo{year}{2009}).

\bibitem[{\citenamefont{Marchukov et~al.}(2016)\citenamefont{Marchukov,
  Volosniev, Valiente, Petrosyan, and Zinner}}]{marchukov2016quantum}
\bibinfo{author}{\bibfnamefont{O.~V.} \bibnamefont{Marchukov}},
  \bibinfo{author}{\bibfnamefont{A.~G.} \bibnamefont{Volosniev}},
  \bibinfo{author}{\bibfnamefont{M.}~\bibnamefont{Valiente}},
  \bibinfo{author}{\bibfnamefont{D.}~\bibnamefont{Petrosyan}},
  \bibnamefont{and} \bibinfo{author}{\bibfnamefont{N.}~\bibnamefont{Zinner}},
  \bibinfo{journal}{Nature communications} \textbf{\bibinfo{volume}{7}},
  \bibinfo{pages}{13070} (\bibinfo{year}{2016}).

\bibitem[{\citenamefont{Cullen and Landau}(1983)}]{cullen1983monte}
\bibinfo{author}{\bibfnamefont{J.~J.} \bibnamefont{Cullen}} \bibnamefont{and}
  \bibinfo{author}{\bibfnamefont{D.}~\bibnamefont{Landau}},
  \bibinfo{journal}{Physical Review B} \textbf{\bibinfo{volume}{27}},
  \bibinfo{pages}{297} (\bibinfo{year}{1983}).

\bibitem[{\citenamefont{Johansson et~al.}(2012)\citenamefont{Johansson, Nation,
  and Nori}}]{johansson2012qutip}
\bibinfo{author}{\bibfnamefont{J.~R.} \bibnamefont{Johansson}},
  \bibinfo{author}{\bibfnamefont{P.~D.} \bibnamefont{Nation}},
  \bibnamefont{and} \bibinfo{author}{\bibfnamefont{F.}~\bibnamefont{Nori}},
  \bibinfo{journal}{Computer physics communications}
  \textbf{\bibinfo{volume}{183}}, \bibinfo{pages}{1760} (\bibinfo{year}{2012}).

\bibitem[{\citenamefont{Rossini et~al.}(2019)\citenamefont{Rossini, Andolina,
  and Polini}}]{rossini2019many}
\bibinfo{author}{\bibfnamefont{D.}~\bibnamefont{Rossini}},
  \bibinfo{author}{\bibfnamefont{G.~M.} \bibnamefont{Andolina}},
  \bibnamefont{and} \bibinfo{author}{\bibfnamefont{M.}~\bibnamefont{Polini}},
  \bibinfo{journal}{Physical Review B} \textbf{\bibinfo{volume}{100}},
  \bibinfo{pages}{115142} (\bibinfo{year}{2019}).

\bibitem[{\citenamefont{Mondal and
  Bhattacharjee}(2022)}]{mondal2022periodically}
\bibinfo{author}{\bibfnamefont{S.}~\bibnamefont{Mondal}} \bibnamefont{and}
  \bibinfo{author}{\bibfnamefont{S.}~\bibnamefont{Bhattacharjee}},
  \bibinfo{journal}{Physical Review E} \textbf{\bibinfo{volume}{105}},
  \bibinfo{pages}{044125} (\bibinfo{year}{2022}).

\bibitem[{\citenamefont{Zhang et~al.}(2019)\citenamefont{Zhang, Yang, Fu, and
  Wang}}]{zhang2019powerful}
\bibinfo{author}{\bibfnamefont{Y.-Y.} \bibnamefont{Zhang}},
  \bibinfo{author}{\bibfnamefont{T.-R.} \bibnamefont{Yang}},
  \bibinfo{author}{\bibfnamefont{L.}~\bibnamefont{Fu}}, \bibnamefont{and}
  \bibinfo{author}{\bibfnamefont{X.}~\bibnamefont{Wang}},
  \bibinfo{journal}{Physical Review E} \textbf{\bibinfo{volume}{99}},
  \bibinfo{pages}{052106} (\bibinfo{year}{2019}).

\bibitem[{\citenamefont{Dziarmaga}(2005)}]{dziarmaga2005dynamics}
\bibinfo{author}{\bibfnamefont{J.}~\bibnamefont{Dziarmaga}},
  \bibinfo{journal}{Physical review letters} \textbf{\bibinfo{volume}{95}},
  \bibinfo{pages}{245701} (\bibinfo{year}{2005}).

\bibitem[{\citenamefont{Caravelli et~al.}(2021)\citenamefont{Caravelli, Yan,
  Garc{\'\i}a-Pintos, and Hamma}}]{caravelli2021energy}
\bibinfo{author}{\bibfnamefont{F.}~\bibnamefont{Caravelli}},
  \bibinfo{author}{\bibfnamefont{B.}~\bibnamefont{Yan}},
  \bibinfo{author}{\bibfnamefont{L.~P.} \bibnamefont{Garc{\'\i}a-Pintos}},
  \bibnamefont{and} \bibinfo{author}{\bibfnamefont{A.}~\bibnamefont{Hamma}},
  \bibinfo{journal}{Quantum} \textbf{\bibinfo{volume}{5}}, \bibinfo{pages}{505}
  (\bibinfo{year}{2021}).

\bibitem[{\citenamefont{B{\"a}uerle et~al.}(2018)\citenamefont{B{\"a}uerle,
  Glattli, Meunier, Portier, Roche, Roulleau, Takada, and
  Waintal}}]{bauerle2018coherent}
\bibinfo{author}{\bibfnamefont{C.}~\bibnamefont{B{\"a}uerle}},
  \bibinfo{author}{\bibfnamefont{D.~C.} \bibnamefont{Glattli}},
  \bibinfo{author}{\bibfnamefont{T.}~\bibnamefont{Meunier}},
  \bibinfo{author}{\bibfnamefont{F.}~\bibnamefont{Portier}},
  \bibinfo{author}{\bibfnamefont{P.}~\bibnamefont{Roche}},
  \bibinfo{author}{\bibfnamefont{P.}~\bibnamefont{Roulleau}},
  \bibinfo{author}{\bibfnamefont{S.}~\bibnamefont{Takada}}, \bibnamefont{and}
  \bibinfo{author}{\bibfnamefont{X.}~\bibnamefont{Waintal}},
  \bibinfo{journal}{Reports on Progress in Physics}
  \textbf{\bibinfo{volume}{81}}, \bibinfo{pages}{056503}
  (\bibinfo{year}{2018}).

\bibitem[{\citenamefont{Mazzoncini et~al.}(2023)\citenamefont{Mazzoncini,
  Cavina, Andolina, Erdman, and Giovannetti}}]{mazzoncini2023optimal}
\bibinfo{author}{\bibfnamefont{F.}~\bibnamefont{Mazzoncini}},
  \bibinfo{author}{\bibfnamefont{V.}~\bibnamefont{Cavina}},
  \bibinfo{author}{\bibfnamefont{G.~M.} \bibnamefont{Andolina}},
  \bibinfo{author}{\bibfnamefont{P.~A.} \bibnamefont{Erdman}},
  \bibnamefont{and}
  \bibinfo{author}{\bibfnamefont{V.}~\bibnamefont{Giovannetti}},
  \bibinfo{journal}{Physical Review A} \textbf{\bibinfo{volume}{107}},
  \bibinfo{pages}{032218} (\bibinfo{year}{2023}).

\bibitem[{\citenamefont{Rubbmark et~al.}(1981)\citenamefont{Rubbmark, Kash,
  Littman, and Kleppner}}]{rubbmark1981dynamical}
\bibinfo{author}{\bibfnamefont{J.~R.} \bibnamefont{Rubbmark}},
  \bibinfo{author}{\bibfnamefont{M.~M.} \bibnamefont{Kash}},
  \bibinfo{author}{\bibfnamefont{M.~G.} \bibnamefont{Littman}},
  \bibnamefont{and} \bibinfo{author}{\bibfnamefont{D.}~\bibnamefont{Kleppner}},
  \bibinfo{journal}{Physical Review A} \textbf{\bibinfo{volume}{23}},
  \bibinfo{pages}{3107} (\bibinfo{year}{1981}).

\bibitem[{\citenamefont{Glasbrenner and
  Schleich}(2023)}]{glasbrenner2023landau}
\bibinfo{author}{\bibfnamefont{E.~P.} \bibnamefont{Glasbrenner}}
  \bibnamefont{and} \bibinfo{author}{\bibfnamefont{W.~P.}
  \bibnamefont{Schleich}}, \bibinfo{journal}{Journal of Physics B: Atomic,
  Molecular and Optical Physics} \textbf{\bibinfo{volume}{56}},
  \bibinfo{pages}{104001} (\bibinfo{year}{2023}).

\bibitem[{\citenamefont{Bose}(2003)}]{bose2003quantum}
\bibinfo{author}{\bibfnamefont{S.}~\bibnamefont{Bose}},
  \bibinfo{journal}{Physical review letters} \textbf{\bibinfo{volume}{91}},
  \bibinfo{pages}{207901} (\bibinfo{year}{2003}).

\bibitem[{\citenamefont{Casanova et~al.}(2012)\citenamefont{Casanova,
  Mezzacapo, Lamata, and Solano}}]{casanova2012quantum}
\bibinfo{author}{\bibfnamefont{J.}~\bibnamefont{Casanova}},
  \bibinfo{author}{\bibfnamefont{A.}~\bibnamefont{Mezzacapo}},
  \bibinfo{author}{\bibfnamefont{L.}~\bibnamefont{Lamata}}, \bibnamefont{and}
  \bibinfo{author}{\bibfnamefont{E.}~\bibnamefont{Solano}},
  \bibinfo{journal}{Physical review letters} \textbf{\bibinfo{volume}{108}},
  \bibinfo{pages}{190502} (\bibinfo{year}{2012}).

\bibitem[{\citenamefont{Zener}(1932)}]{zener1932non}
\bibinfo{author}{\bibfnamefont{C.}~\bibnamefont{Zener}},
  \bibinfo{journal}{Proceedings of the Royal Society of London. Series A,
  Containing Papers of a Mathematical and Physical Character}
  \textbf{\bibinfo{volume}{137}}, \bibinfo{pages}{696} (\bibinfo{year}{1932}).

\bibitem[{\citenamefont{Landau}(1932)}]{landau1932theorie}
\bibinfo{author}{\bibfnamefont{L.}~\bibnamefont{Landau}},
  \bibinfo{journal}{Physikalische Zeitschrift der Sowjetunion}
  \textbf{\bibinfo{volume}{2}}, \bibinfo{pages}{46} (\bibinfo{year}{1932}).

\bibitem[{\citenamefont{Wang et~al.}(2008)\citenamefont{Wang, Huang, and
  Yi}}]{wang2008landau}
\bibinfo{author}{\bibfnamefont{L.}~\bibnamefont{Wang}},
  \bibinfo{author}{\bibfnamefont{X.}~\bibnamefont{Huang}}, \bibnamefont{and}
  \bibinfo{author}{\bibfnamefont{X.}~\bibnamefont{Yi}}, \bibinfo{journal}{The
  European Physical Journal D} \textbf{\bibinfo{volume}{46}},
  \bibinfo{pages}{345} (\bibinfo{year}{2008}).

\bibitem[{\citenamefont{Guo et~al.}(2024)\citenamefont{Guo, Yang, and
  Dou}}]{guo2024analytically}
\bibinfo{author}{\bibfnamefont{W.-X.} \bibnamefont{Guo}},
  \bibinfo{author}{\bibfnamefont{F.-M.} \bibnamefont{Yang}}, \bibnamefont{and}
  \bibinfo{author}{\bibfnamefont{F.-Q.} \bibnamefont{Dou}},
  \bibinfo{journal}{Physical Review A} \textbf{\bibinfo{volume}{109}},
  \bibinfo{pages}{032201} (\bibinfo{year}{2024}).

\bibitem[{\citenamefont{Polkovnikov et~al.}(2011)\citenamefont{Polkovnikov,
  Sengupta, Silva, and Vengalattore}}]{polkovnikov2011colloquium}
\bibinfo{author}{\bibfnamefont{A.}~\bibnamefont{Polkovnikov}},
  \bibinfo{author}{\bibfnamefont{K.}~\bibnamefont{Sengupta}},
  \bibinfo{author}{\bibfnamefont{A.}~\bibnamefont{Silva}}, \bibnamefont{and}
  \bibinfo{author}{\bibfnamefont{M.}~\bibnamefont{Vengalattore}},
  \bibinfo{journal}{Reviews of Modern Physics} \textbf{\bibinfo{volume}{83}},
  \bibinfo{pages}{863} (\bibinfo{year}{2011}).

\bibitem[{\citenamefont{Heyl et~al.}(2013)\citenamefont{Heyl, Polkovnikov, and
  Kehrein}}]{heyl2013dynamical}
\bibinfo{author}{\bibfnamefont{M.}~\bibnamefont{Heyl}},
  \bibinfo{author}{\bibfnamefont{A.}~\bibnamefont{Polkovnikov}},
  \bibnamefont{and} \bibinfo{author}{\bibfnamefont{S.}~\bibnamefont{Kehrein}},
  \bibinfo{journal}{Physical review letters} \textbf{\bibinfo{volume}{110}},
  \bibinfo{pages}{135704} (\bibinfo{year}{2013}).

\end{thebibliography}

\end{document}